\documentclass[]{JHEP3} 

\usepackage{graphicx,epsfig,multicol,bbm,epsfig,amsmath}

\newcommand\fverb{\setbox\fverbbox=\hbox\bgroup\verb}
\newcommand\fverbdo{\egroup\medskip\noindent%
			\fbox{\unhbox\fverbbox}\ }
\newcommand\fverbit{\egroup\item[\fbox{\unhbox\fverbbox}]}
\newbox\fverbbox

\title{Dependence of direct detection signals on the WIMP velocity distribution} 
\author{Anne M.~Green\\
School of
Physics and Astronomy, University of Nottingham, University Park,
  Nottingham, NG7 2RD, UK\\
E-mail: \email{anne.green@nottingham.ac.uk}}

\abstract{ 
  The signals expected in WIMP direct detection experiments depend on
  the ultra-local dark matter distribution. Observations probe the local
  density, circular speed and escape speed, while simulations
find velocity distributions that deviate significantly from the standard
   Maxwellian distribution.  We calculate the energy, time and direction
  dependence of the event rate for a range of velocity distributions
  motivated by recent observations and simulations, and also investigate the uncertainty
  in the determination of WIMP parameters.  The dominant uncertainties
  are the systematic error in the local circular
  speed and whether or not  the MW has a high density dark disc.
 In both cases there are substantial changes in the mean differential
 event rate and the annual modulation signal, and hence exclusion
 limits and determinations of the WIMP mass.  The uncertainty in the
 shape of the halo velocity distribution is less important, however
 it leads to a $\sim 5\%$ systematic error in the WIMP mass.
The detailed direction dependence of the event rate is sensitive
to the velocity distribution. However the numbers of events required to 
detect anisotropy and confirm the median recoil direction do not change substantially.} 
 

\preprint{}
 
\begin{document}

\section{Introduction} 
\label{intro} 
 
Weakly Interacting Massive Particles (WIMPs) are a well motivated
dark matter candidate~\cite{jkg,bhs,dmbook}.  They can be directly
detected in the lab via their elastic scattering off target nuclei
in dedicated detectors~\cite{dd} and experiments are now probing the
theoretically favoured regions of parameter
space~\cite{zeplin,cdms,xenon100}.

The nuclear recoil event rate is energy, time and direction dependent.
Due to the Earth's orbit about
the Sun the net velocity of the lab with respect to the Galactic rest
frame varies over the course of a year. The net speed is largest in
the Summer and hence there are more high speed WIMPs, and less low
speed WIMPs, in the lab frame. This produces an energy dependent
annual modulation in the event rate with amplitude of order a few
per-cent~\cite{dfs}. The WIMP flux in the lab frame is sharply peaked
in the direction of motion of the Sun (towards the constellation
CYGNUS), and hence the recoil spectrum is peaked in the direction
opposite to this~\cite{spergel}. This directional signal is far larger than the
annual modulation; the event rate in the backward direction is roughly
an order of magnitude larger than that in the forward direction~\cite{spergel}. A
detector which can measure the recoil directions is required to detect
this signal  (see Ref.~\cite{ahlen} for an overview of the status of
directional detection experiments).
The time and direction dependence of the event rate are signals
which can be used to discriminate WIMP induced recoils from
backgrounds, while the energy dependence of the event rate can be used
to measure the WIMP mass~\cite{ls,bk,brown,mpap1,sd,mpap2}.

The energy~\cite{kk,donato},
time~\cite{br,belli,vergados,ullio,ecz,am1,am2,nata} and
direction~\cite{ck1,ck2,morganpap1} dependence of the event rate all
depend on the ultra-local WIMP distribution.  Historically analytic halo
models have been used to calculate the WIMP signals and analyse data.
In the past few years velocity distributions from high resolution
simulations of the formation of Milky Way like
halos~\cite{hansen,read1,vogelsberger,read2,kuhlen,ling1} have been
used to calculate some of the WIMP
signals~\cite{vhh,bruch,fs,vogelsberger,mmm,kuhlen,ling1,ling2,mccabe}.

In this paper we study the energy, time
and direction dependence of the elastic scattering 
event rate for a range of velocity distributions motivated by recent observations and
simulations. Where possible we use WIMP and
target nuclei independent parameterisations of the event rate. We
also investigate the uncertainty in exclusion limits,
determinations of the WIMP mass and the number of events required to
detect the directional signal.  For
studies of the impact of uncertainties in the WIMP velocity
distribution on the interpretation of data from recent experiments see
Refs.~\cite{fs,kuhlen,ling1,ling2,mccabe,cbc} for elastic scattering,
Refs.~\cite{mmm,kuhlen,ling1,mccabe} for inelastic scattering and
Ref.~\cite{mccabe} for momentum dependent scattering.

In Sec.~\ref{sig} we review the calculation of the event
rate. In Sec.~\ref{dist} we discuss recent results, in particular from
numerical simulations, on the local dark matter distribution and
present the velocity distributions which we will use to calculate the
WIMP signals in Sec.~\ref{ddsig}. Finally we conclude with a summary
in Sec.~\ref{discuss}.

\section{Event rate}
\label{sig}

Assuming spin-independent coupling the differential event rate (number
of events per unit energy, per unit time, per unit detector mass) is
given by~\cite{jkg,ls}
\begin{equation}
\label{drde}
\frac{{\rm d} R}{{\rm d}E}(E,t) =
             \frac{\sigma_{{\rm p}} 
             \rho_{\chi}}{2 \mu_{{\rm p} \chi}^2 m_{\chi}}
             A^2 F^2(E)   \int^{\infty}_{v_{{\rm min}}} 
            \frac{f(v,t)}{v} {\rm d}v     \,, 
\end{equation}
where $\rho_{\chi}$ is the ultra-local WIMP density, $f(v, t)$ the normalised
ultra-local WIMP speed distribution in the rest frame of the detector,
$\sigma_{{\rm p}}$ the WIMP scattering cross section on the proton,
$\mu_{{\rm p} \chi} = (m_{\rm p} m_{\chi})/(m_{{\rm p}}+ m_{{\chi}})$
the WIMP-proton reduced mass, $A$ and $F(E)$ the mass number and form
factor of the target nuclei respectively and $E$ is the recoil energy.
The lower limit of the integral, $v_{{\rm min}}$, is the minimum WIMP
speed that can cause a recoil of energy $E$:
\begin{equation}
\label{vmin}
v_{{\rm min}}=\left( \frac{ E m_{A}}{2 \mu_{{\rm A} \chi}^2} 
             \right)^{1/2} \,,
\end{equation}
where $m_{A}$ is the atomic mass of the detector nuclei
and $\mu_{{\rm A} \chi}$ the WIMP-nucleon reduced mass.

The WIMP speed distribution in the detector rest frame is calculated
by carrying out, a time dependent, Galilean transformation: ${\bf v}
\rightarrow \tilde{\bf{v}} = {\bf v} + {\bf
  v}_{e}(t)$~\footnote{Formally the effects of gravitational focusing
  by the Sun should be taken into account~\cite{griest,ag2}, however
  the resulting modulation in the differential event rate is small and
  only detectable with a very large number of events~\cite{griest}.}.  The Earth's
motion relative to the Galactic rest frame, ${\bf v}_{e}(t)$, is made
up of three components: the motion of the Local Standard of Rest
(LSR), ${\bf v}_{\rm LSR}$, the Sun's peculiar motion with respect to the LSR,
${\bf v}_{\odot}^{\rm p}$, and the Earth's orbit about the Sun, ${\bf
  v}_{e}^{\rm orb}$.  We use ${\bf v}_{\rm LSR}=(0, v_{\rm c}, 0)$,
where $v_{\rm c}$ is the local circular speed,  and $ {\bf v}_{\odot}^{\rm p} = (11.1, \,
12.2, 7.3) \, {\rm km \, s}^{-1}$~\cite{schoenrich} in Galactic
co-ordinates (where $x$ points towards the Galactic center, $y$ is the
direction of Galactic rotation and $z$ towards the North Galactic
Pole) and the expressions for the Earth's orbit from
Ref.~\cite{book}.

The differential event rate, eq.~(\ref{drde}), depends on the target nuclei mass and also the (a priori unknown) WIMP mass. It is therefore useful (c.f. Ref.~\cite{kk}) to consider the `WIMP and target independent' quantity
\begin{equation}
\label{t}
T(v_{\rm min},t )   = (220 \, {\rm km \, s}^{-1}) \int_{v_{\rm min}}^{\infty} \frac{f(v,t)}{v} \, {\rm d} v \,.
\end{equation}
The prefactor here is chosen to make $T(v_{\rm min}, t)$ dimensionless
(and of order unity) while allowing variations in the local circular
speed (see Sec.~\ref{secvc}). The differential event rate can then be
written as
\begin{equation}
\frac{{\rm d} R}{{\rm d}E}(E,t) = \frac{C({\chi}, {\rm A})
  \rho_{\chi}}{220 \, {\rm km \, s}^{-1}} \, T(v_{\rm min},t ) \,,
\end{equation}
where the prefactor
\begin{equation}
\label{C}
C({\chi}, {\rm A})= \frac{\sigma_{{\rm p}}}{2 \mu_{{\rm p} \chi}^2 m_{\chi}}
             A^2 F^2(E) \,,
\end{equation}
contains the WIMP and target dependent terms and is independent of the
astrophysical WIMP distribution.

The direction dependence~\cite{spergel} of the event rate is most
compactly written in terms of the radon transform of the WIMP velocity
distribution~\cite{gondolo}
\begin{equation}
\frac{{\rm d}R}{{\rm d} E \, {\rm d} \Omega} = 
\frac{\rho_{\chi} \sigma_{\rm p} A^2}{4 \pi \mu_{{\rm p} \chi}^2 m_{\chi}} F^2(E) 
   \hat{f}(v_{\rm min},\hat{{\bf q}}) \,,
\end{equation}
where ${\rm d} \Omega = {\rm d} \phi \, {\rm d} \cos{\gamma}$,
$\hat{\bf{q}}$ is the recoil direction and
$\hat{f}(v_{\rm min},\hat{{\bf q}})$ is the 3-dimensional Radon transform
of the WIMP velocity distribution $f({\bf v})$
\begin{equation}
\hat{f}(v_{\rm min},\hat{{\bf q}}) = \int \delta({\bf v}.\hat{{\bf q}} - v_{\rm min}) f({\bf v})
 {\rm d}^3 v \,.
\end{equation}
Geometrically the Radon transform, $\hat{f}(v_{\rm min},\hat{{\bf q}})$,
is the integral of the function $f({\bf v})$ on a plane
orthogonal to the direction $\hat{\bf{q}}$ at a distance $v_{\rm min}$ from the origin. See Ref.~\cite{ck2} for an alternative, but equivalent, expression.

While the directional recoil rate depends on both of the angles which
specify a given direction, the strongest signal is the differential of
the event rate with
respect to the angle between the recoil and the direction of solar
motion, $\gamma$,~\cite{spergel,ag}
\begin{equation}
\label{drdedcos}
\frac{{\rm d}^2 R}{{\rm d} E \, {\rm d}\cos{\gamma}} 
  = \frac{\rho_{\chi} \sigma_{\rm p}}{4 \pi \mu_{{\rm p}\chi} m_{\chi}} A^2 F^2(E) 
  \int_{0}^{2 \pi}   \hat{f}(v_{\rm min},\hat{{\bf q}}) \, {\rm d} \phi \,,
\end{equation}
where $\phi$ is the azimuthal angle. As with the non-directional
differential event it is useful to consider a dimensionless `WIMP and target
independent' quantity
\begin{equation}
\label{tdir}
{\cal T}(v_{\rm min}, \cos{\gamma}) = (220 \, {\rm km \, s}^{-1}) \int_{0}^{2 \pi}   \hat{f}(v_{\rm min},\hat{{\bf q}}) \, {\rm d} \phi \,,
\end{equation}
so that eq.~(\ref{drdedcos}) can be rewritten as 
\begin{equation}
\frac{{\rm d}^2 R}{{\rm d} E \,{\rm d}\cos{\gamma}} 
  =
\frac{C({\chi}, {\rm A}) \rho_{\chi}}{2 \pi (220 \, {\rm km \,
    s}^{-1})} \,
{\cal T}(v_{\rm min}, \cos{\gamma}) \,,
\end{equation}
where $C(\chi, {\rm A})$ is defined in eq.~(\ref{C}).

\section{Dark matter distribution} 
\label{dist}

The traditional benchmark model for direct detection event rate
calculations is the standard halo model (an isotropic isothermal
sphere with $\rho(r) \propto r^{-2}$)  which has velocity
distribution
\begin{equation}
f({\bf v}) \propto
\begin{cases}
 \exp{\left( - \frac{|{\bf v}|^2}{2 \sigma^2} \right)} & \text{if $|{\bf v}| < v_{\rm esc}$\,} \\
 0 & \text{otherwise} \,.
 \end{cases}
\end{equation}
where $\sigma$ is the velocity dispersion. 

In Sec.~\ref{obs} we review the status of observational
measurements of the local dark matter density, circular speed and
escape speed, and their relevance to the event rate calculation.
In Sec.~\ref{simfv} we review results on the shape of the dark matter
velocity distribution from recent numerical simulations. We conclude
in Sec.~\ref{modelsum} with a summary of the benchmark models
which we will use to calculate the direct detection signals in
Sec.~\ref{ddsig}.

\subsection{Observations}
\label{obs}

\subsubsection{Local density}
The local density appears in the normalisation of the event rate. We
use the standard value,  $\rho_{\chi} \sim 0.3 \, {\rm GeV} \, {\rm
  cm}^{-3}$. For a given model for the MW it is possible to determine
$\rho_{\chi}$ to $\sim 10\%$ accuracy~\cite{widrow,cu}, however analyses
which use a range of models or model independent methods find
substantially larger, of order unity,
uncertainties~\cite{weber,garbari,salucci}.  Furthermore observations determine
the local density averaged over a spherical shell, and the DM density in
the stellar disc of simulated halos is $\sim 20 \%$ larger than the
shell average~\cite{pato}.

Since the normalisation is proportional to the product of $\rho_{\chi}$
and $\sigma_{\rm p}$, the uncertainty in $\rho_{\chi}$ translates
directly into an uncertainty in constraints on, or measurements of,
$\sigma_{\rm p}$. As this uncertainty is the same for all experiments
we do not explicitly consider it.

\subsubsection{Local circular speed}
\label{secvc}

The local circular speed, the speed with which stars orbit the Galactic centre at the solar radius,
is related to the velocity dispersion by the Jean's equation~\cite{bt}.
\begin{equation}
 \frac{1}{\rho} \frac{{\rm d} (\rho \sigma_{r}^2)}{{\rm d} r} + 2 \frac{\beta \sigma_{r}^2}{r}  =- \frac{v_{\rm c}^2}{r} \,,
\end{equation}
where $\sigma_{\rm r}$ is the radial velocity dispersion and $\beta =
1 - (\sigma_{\theta}^2 + \sigma_{\phi}^2)/2 \sigma_{r}^2$ is the
anisotropy parameter.  For the standard halo model the velocity dispersion is  isotropic
($\sigma_{r}=\sigma_{\theta} = \sigma_{\phi}$ so that $\beta=0$) and
independent of radius and $\rho(r) \propto r^{-2}$ so that
$\sigma_{r}= v_{\rm c}/\sqrt{2}$.

The standard value of $v_{\rm c}$ is $v_{c} = 220 \, {\rm km \,
  s}^{-1}$~\cite{kerr}. A recent analysis using Galactic
masers found a significantly higher value, $v_{\rm c} = (254 \pm 16)
\, {\rm km \, s}^{-1}$,~\cite{reid} and this has been adopted is some subsequent
direct detection work~\cite{savagevc}.  It has been argued, however, that this
analysis used overly restrictive models.  Bovy et al. found $v_{\rm c}
= (236 \pm 11) {\rm km \, s}^{-1}$, assuming a flat rotation
curve~\cite{bovy}, while McMillan and Binney find values ranging from
$v_{\rm c} = (200 \pm 20) {\rm km \, s}^{-1}$ to $v_{\rm c} = (279 \pm
33) {\rm km \, s}^{-1}$ depending on the model used for the rotation
curve~\cite{mcmillan}.

Given the significant systematic uncertainties in determinations of
$v_{\rm c}$, we retain $v_{\rm c}=220 \, {\rm km \, s}^{-1}$ as the
default value, and also consider $v_{\rm c}=200$ and $280 \, {\rm km \,
  s}^{-1}$. 

\subsubsection{Local escape speed}

Particles with speed, in the Galactic rest frame greater than the
local escape speed, $v_{esc}= \sqrt{2 |\Phi(R_{0})|}$ where $\Phi(r)$
is the potential, are not gravitationally bound. The standard halo
model formally extends to infinity and therefore the speed
distribution has to be truncated at $v_{esc}$ `by hand' (see
e.g. Ref.~\cite{dfs}).  Historically the standard value for the escape speed was
$v_{esc} = 650 \, {\rm km \, s}^{-1}$.  A more recent analysis, using
high velocity stars from the RAVE survey, finds $498 \, {\rm km \,
  s}^{-1} < v_{esc} < 608 \, {\rm km \, s}^{-1}$ with a median likelihood
of $544 \, {\rm km \, s}^{-1}$~\cite{smith}.

We use as a default the median RAVE value, $v_{\rm esc} =544 \, {\rm km \, s}^{-1}$, and also
consider the old standard value, $v_{\rm esc} =650 \, {\rm km \, s}^{-1}$.

\subsection{Simulations}
\label{simfv}

High resolution dark matter only simulations of the formation of Milky Way like dark
matter halos,  find speed distributions which deviate
systematically from a multivariate Gaussian (the simplest anisotropic
generalisation of the Maxwellian
distribution)~\cite{hansen,fs,vogelsberger,kuhlen}. There are more low
speed particles, and the peak in the distribution is lower. The
deviation is smaller in the lab frame than in the Milky Way
rest frame however~\cite{kuhlen}. Kuhlen et
al.~\cite{kuhlen} study the velocity distributions of the particles in
the Via Lactea 2 (VL2)~\cite{VL2} and GHALO~\cite{GHALO}
simulations. In each case they consider the particles centered in a 1
kpc shell centered on galactocentric radius $r=8.5 \, {\rm kpc}$ and
also 100 sample spheres of radii 1 or 1.5 kpc, each centered at a point with $r=8.5
\, {\rm kpc}$. The shell contains a large number ($\sim 10^{6}$)
particles, allowing the average velocity distribution at the solar
radius to be measured with small statistical errors. The spheres
contain a smaller number of particles ($\sim 10^{4}$), and hence have
larger statistical errors, but are sensitive to local variations in the
velocity distribution on $\sim \, {\rm kpc}$ scales.
They find that the radial, $v_{\rm r}$ and tangential, $v_{\rm
  t}=\sqrt{v_{\theta}^2 + v_{\phi}^2}$, velocity distributions are well
fit, apart from at large $v_{\rm t}$, by modified gaussian
distributions~\cite{fs}:
\begin{eqnarray}
\label{fvsim}
f(v_{\rm r}) &=& \frac{1}{N_{\rm r}} \exp{ \left[ - \left( \frac{v_{\rm r}}{\bar{v}_{\rm r}} \right)^{\alpha_{\rm r}} \right]} \,, \\
f(v_{\rm t}) &=& \frac{v_{\rm t}}{N_{\rm t}} \exp{ \left[ - \left( \frac{v_{\rm t}}{\bar{v}_{\rm t}} \right)^{\alpha_{\rm t}} \right]} \,,
\end{eqnarray}
where $N_{\rm r/t}$ are normalisation factors. 
The velocity distributions also have stochastic features at high speeds. There are broad
bumps which vary from halo to halo, but are independent of position
within a given halo and are thought to reflect the formation history
of the halo~\cite{vogelsberger,kuhlen}. Kuhlen et al.~\cite{kuhlen}
also find narrow spikes in some locations, corresponding to tidal
streams.

We use the shell and sphere median, 16th and 84th percentile fit
parameters for the VL2~\cite{VL2} simulation, given in Table 1 of
Ref.~\cite{kuhlen}.  As discussed by Kuhlen et al., the velocity
dispersion and circular speed of these dark matter only simulations is
lower than expected in reality; baryonic contraction will deepen the
potential well and increase the circular speed. For VL2 the most
likely speed is $v_{0}=184 \, {\rm km \, s}^{-1}$ and $v_{\rm c}/v_{0}
\approx 0.85$. To allow a comparison with the standard halo model
(with standard parameters) we scale the fit parameters so that the
peak of the speed distribution matches that of the standard halo
model. We truncate the fitted velocity distributions `by hand' at the
median value from RAVE, $v_{\rm vesc}= 544 \, {\rm km \, s}^{-1}$. For
the time averaged differential event rate we also consider the
tabulated data from the VL2 simulation from Ref.~\cite{kuhlenweb}. As
discussed by Ref.~\cite{kuhlen} in this case the scaling results in
some of the high speed features in the distribution being pushed
beyond the escape speed. These features are likely to be significant
for experiments which are only sensitive to the high speed tail of the
speed distribution.

It should be cautioned that the scales resolved by simulations are
many orders of magnitude larger than those probed by direct detection
experiments.  Vogelsberger and White have developed a new technique to
study the ultra-local dark matter distribution~\cite{vw}. They find
that the ultra-local dark matter distribution consists of a huge
number of streams and is essentially smooth. Schneider et
al.~\cite{skm} have reached similar conclusions by studying the
evolution of the first, Earth mass, microhalos to
form~\cite{hss,ghs,dms}. This suggests (see also Ref.~\cite{kk2}) that
the ultra-local dark matter density and velocity distribution should
not be drastically different (i.e. composed of a small number of
streams) to those on the scales resolved by simulations. The
ultra-local velocity distribution may, however, contain some
fine-grained substructure~\cite{fantin,afshordi}.

The simulations discussed above contain dark matter only, while
baryons dominate in the inner regions of the Milky Way. Simulating
baryonic physics is extremely difficult, and producing galaxies whose
detailed properties match those of real galaxies is an outstanding
challenge. Some recent simulations have found that late merging
sub-halos are preferentially dragged towards the disc, where they are
destroyed leading to the formation of a co-rotating dark
disc (DD)~\cite{read1,read2,ling1}. 

The properties (density and velocity distribution) of the DD
 are highly uncertain. We consider
3 benchmark models, which aim to broadly span the range of
plausible properties. The first benchmark model follows
Ref.~\cite{bruch} modelling the DD velocity distribution as a 
gaussian with isotropic dispersion, $\sigma_{\rm DD}=50 \, {\rm km \,
  s}^{-1}$ and lag $v_{\rm lag} = 50\, {\rm km \, s}^{-1}$, matching
(roughly) the kinematics of the Milky Way's stellar thick disc. We use
the central value of the DD density from Ref.~\cite{bruch},
$\rho_{\rm DD}= \rho_{\rm H}$, where $\rho_{\rm H}$ is the local halo
density.

Ref.~\cite{purcell} argues that to be consistent with the observed
morphological and kinematic properties of the Milky Way's thick disc,
the Milky Way's merger history must be quiescent compared with typical
$\Lambda$CDM merger histories. Hence the DD density must be relatively
small, $\rho_{\rm DD} < 0.2 \rho_{\rm H}$, at the lower end of the
range of values considered in
Ref.~\cite{bruch}. Refs.~\cite{purcell,ling2} also argue that the DD
velocity dispersion is likely to be substantially larger than that of
the stellar thick disc. In order to study the effects of increasing
the DD velocity dispersion and decreasing the density we consider two
further benchmark models, one with $\sigma_{\rm DD}=50 \, {\rm km \,
  s}^{-1}$ and $\rho_{\rm DD}= 0.15 \rho_{\rm H}$ and one with
$\sigma_{\rm DD}=100 \, {\rm km \, s}^{-1}$ and $\rho_{\rm DD}= 0.15
\rho_{\rm H}$. In both cases we keep $v_{\rm lag} = 50 \, {\rm km \,
  s}^{-1}$ and maintain the assumption of an isotropic gaussian
velocity distribution. Ref.~\cite{ling2} argues that the DD velocity
distribution is better fit by a Tsallis
distribution~\cite{tsallis}. Given the substantial uncertainties in
the DD density and velocity dispersion we do not investigate the
effect of the uncertainty in the shape of the dark disc
velocity distribution. For all three benchmark dark disc models, for
simplicity and following Ref.~\cite{bruch}, we use the standard halo
model with standard parameters for the dark matter halo and fix the
total local density to the standard value $\rho_{\rm H} + \rho_{\rm
  DD} = 0.3 \, {\rm GeV} \, {\rm cm}^{-3}$.

\TABULAR[ht]{||c||l||}{
  \hline
  label & properties \\ \hline
  \hline
  SHSP & standard halo model (SH) with $v_{\rm c}= \, 220  \, {\rm km
    \, s}^{-1}$, $v_{\rm esc} =\,554 \, {\rm km \, s}^{-1}$  \\ \hline
  SH$v_{\rm esc}$H & SH with $v_{\rm c}=\, 220  {\rm km \, s}^{-1}$, $v_{\rm esc} =650\, {\rm km \, s}^{-1}$  \\ \hline
  SH$v_{\rm c}$L & SH with $v_{\rm c}= 200\,  {\rm km \, s}^{-1}$, $v_{\rm esc} =554 \,{\rm km \, s}^{-1}$  \\ \hline
  SH$v_{\rm c}$H  & SH with $v_{\rm c}= 280\,  {\rm km \, s}^{-1}$, $v_{\rm esc} =554 \,{\rm km \, s}^{-1}$  \\\hline \hline
  SIMsh &  scaled modified Maxwellian fit to VL2 shell data,
  eq.~(\ref{fvsim}) from Ref.~\cite{kuhlen}: \\
 & 
  $\bar{v}_{\rm r} = 1.2 \times 202 \,{\rm km \, s}^{-1}$,
  $\bar{v}_{\rm t} = 1.2 \times 129 \,{\rm km \, s}^{-1}$, $\alpha_{\rm r}= 0.93$, $\alpha_{\rm t}=0.64$.
  \\ \hline
 SIMspmed & scaled median modified Maxwellian fit to VL2 sphere
 data:\\ & 
$\bar{v}_{\rm r} = 1.2 \times 200 \,{\rm km \, s}^{-1}$,
  $\bar{v}_{\rm t} = 1.2 \times 135 \,{\rm km \, s}^{-1}$, $\alpha_{\rm r}= 0.94$, $\alpha_{\rm t}=0.66$.
\\ \hline
SIMsp16 &scaled 16th percentiles modified Maxwellian fit
to VL2 sphere data: \\ & 
$\bar{v}_{\rm r} = 1.2 \times 186 \,{\rm km \, s}^{-1}$,
  $\bar{v}_{\rm t} = 1.2 \times 124 \,{\rm km \, s}^{-1}$, $\alpha_{\rm r}= 0.88$, $\alpha_{\rm t}=0.64$.
  \\ \hline
SIMsp84 & scaled 84th percentiles modified Maxwellian fit to VL2
sphere data:\\ & 
$\bar{v}_{\rm r} = 1.2 \times 213 \,{\rm km \, s}^{-1}$,
  $\bar{v}_{\rm t} = 1.2 \times 149 \,{\rm km \, s}^{-1}$, $\alpha_{\rm r}= 0.99$, $\alpha_{\rm t}=0.67$.
\\ \hline  \hline
DD$\rho$H$\sigma$L & SHSP plus a dark disk with $v_{\rm lag}= 50\, {\rm km \, s}^{-1}$,
$\rho_{\rm DD} = \rho_{\rm H}$, $\sigma_{\rm DD}= 50 \,{\rm km \, s}^{-1}$    \\ \hline
DD$\rho$L$\sigma$L & SHSP plus a dark disk with $v_{\rm lag}= 50\, {\rm km \, s}^{-1}$,
$\rho_{\rm DD} = 0.15 \rho_{\rm H}$, $\sigma_{\rm DD}= \,50 {\rm km \, s}^{-1}$  \\ \hline
DD$\rho$L$\sigma$H &  SHSP plus a dark disk with $v_{\rm lag}= 50 \,{\rm km \, s}^{-1}$,
$\rho_{\rm DD} = 0.15 \rho_{\rm H}$, $\sigma_{\rm DD}= 100 \,{\rm km \, s}^{-1}$   \\ \hline
\hline
\hline}
{\label{models} Summary of benchmark models.}

\subsection{Summary}
\label{modelsum}

The benchmark models we consider are summarised in Table 1.
For compactness we refer to the
standard halo model with $v_{\rm c} = 220 \, {\rm km \, s}^{-1}$ and
$v_{\rm esc} = 544 \, {\rm km \, s}^{-1}$ as the standard halo model
with standard parameters (SHSP).  The normalised speed distribution for each
of the models is plotted in fig.~\ref{fig-fv}.  For the standard halo
model, increasing (decreasing) $v_{\rm c}$ increases (decreases) both
$v_{0}$, the value of $v$ at which $f(v)$ peaks, and also the width of
$f(v)$. The speed distribution of a pure DD has the same qualitative
shape as the standard halo model. The effect of a DD on the total
normalised speed distribution depends strongly on both the DD
density and speed dispersion. With a high DD density and a low
DD velocity dispersion there is a large additional peak in the speed
distribution as low speed. As the DD density is decreased the
height of the peak decreases. If the DD velocity dispersion is
increased the separation of the DD speed peak and halo speed
peak decreases. For DD$\rho$L$\sigma$H, which has a low DD
density and a large speed dispersion, the speed distribution has a
single peak, at a lower speed than the standard halo model.  As
previously found~\cite{kuhlen} the speed distributions from the
modified Maxwellian fits to the VL2 simulation data have less low
speed and more high speed particles than the standard halo model with
the same peak speed, $v_{0}$. However the differences, in the lab
frame, are fairly small~\cite{kuhlen}. The best fit to the shell data and the median
fit to the sphere data are fairly similar.  The scatter between sphere
fits is ${\cal O}(10 \%)$. If the simulation fits were compared to a
standard halo with the same circular speed, $v_{\rm c}$ (which is the
observable quantity) rather than the same peak speed, $v_{0}$, the
deviations from the standard halo model would, however, be larger.

\FIGURE[ht]{
\epsfig{file=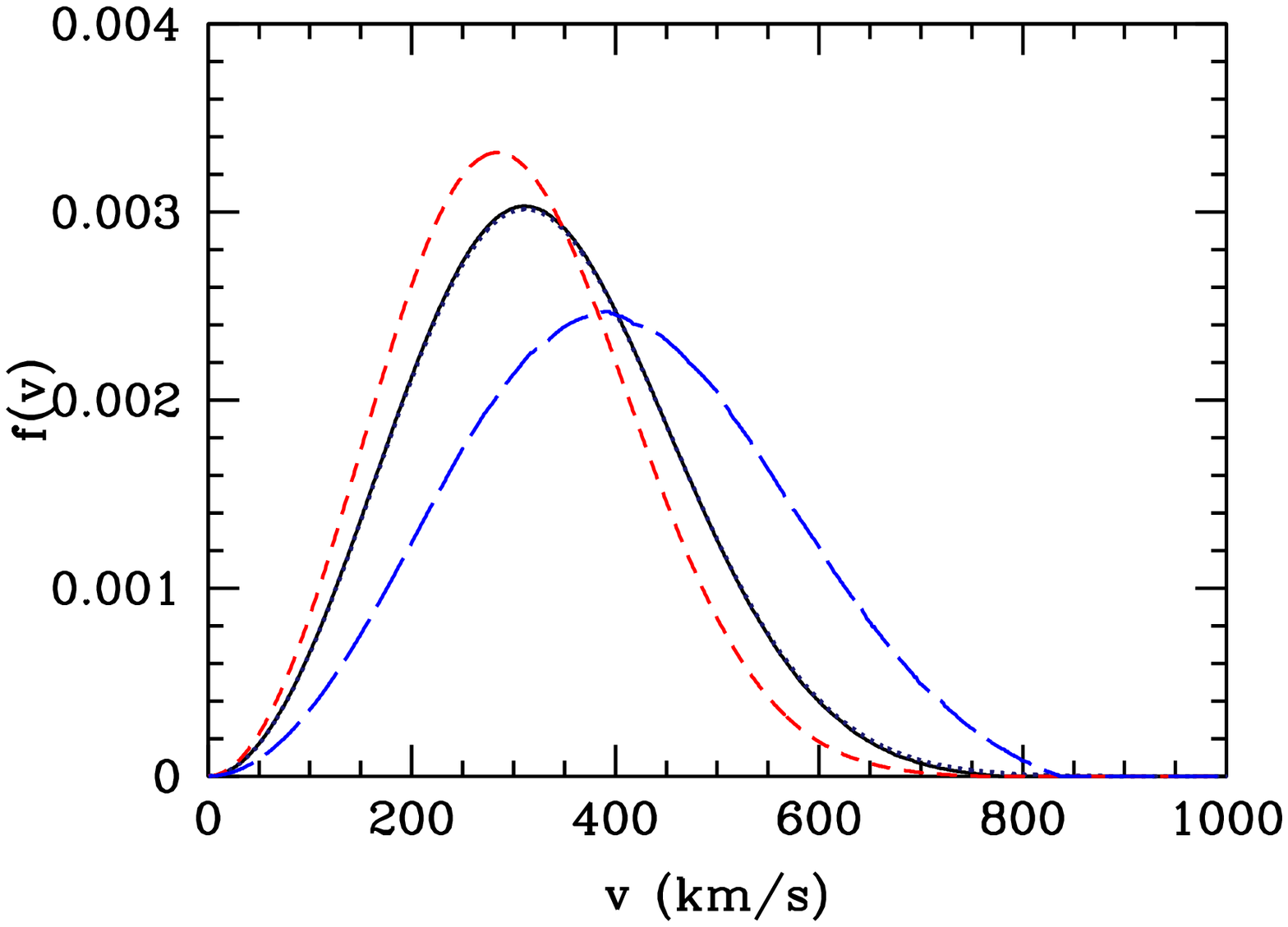,width=7.0cm}\\
\epsfig{file=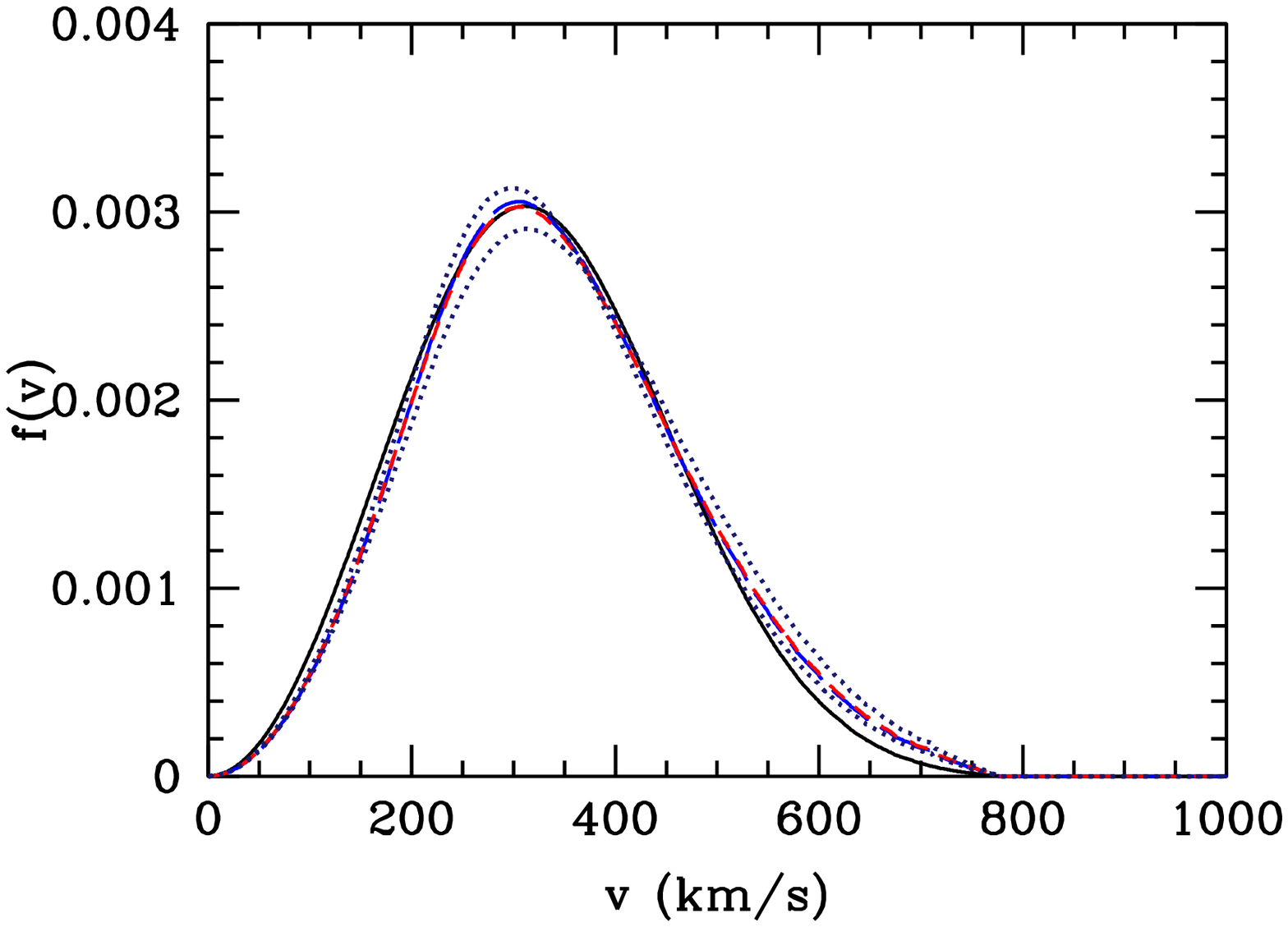,width=7.0cm}
\epsfig{file=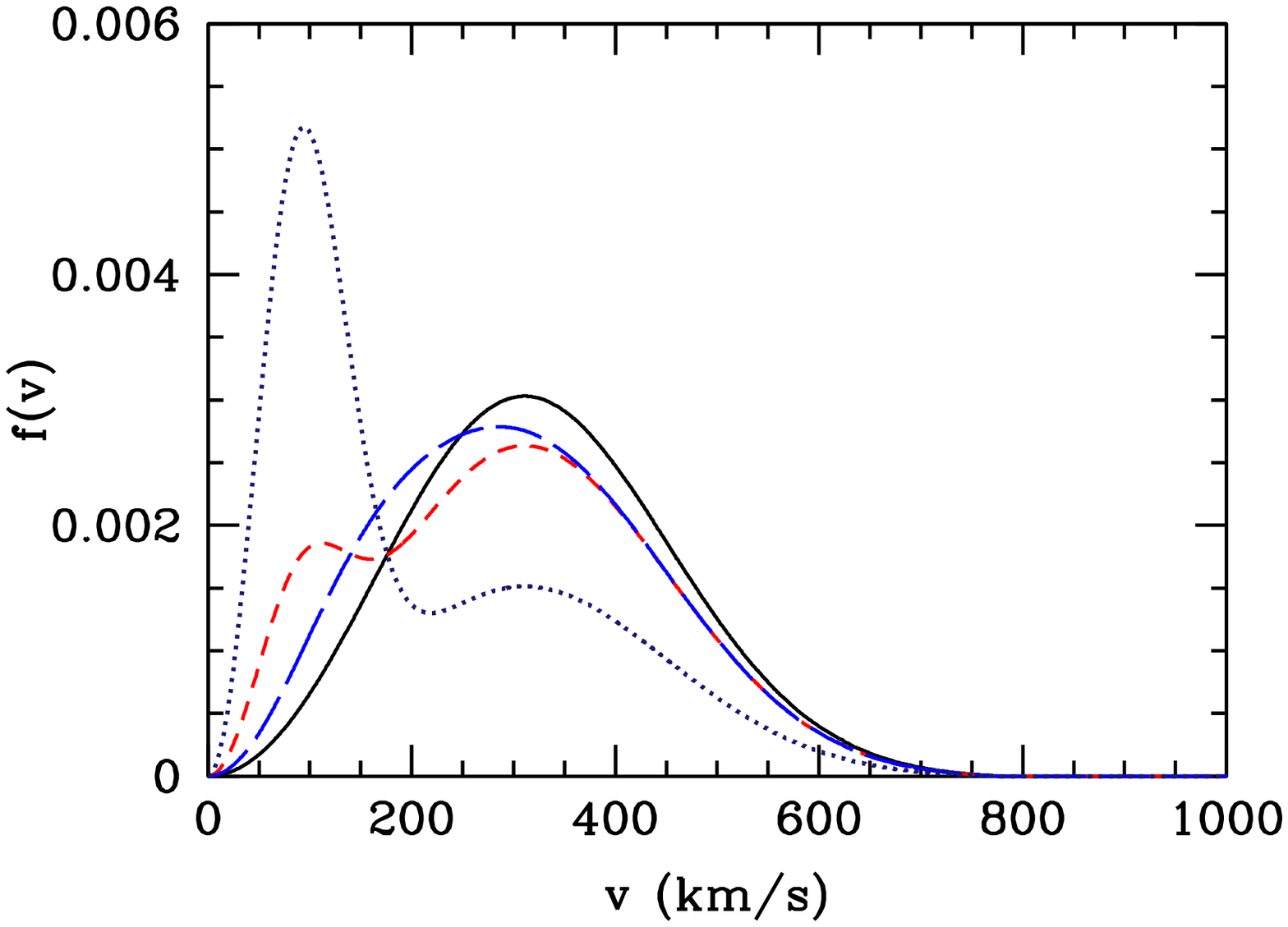,width=7.0cm}
\caption{The normalised speed distributions.  Top panel, the standard
  halo model: SHSP (solid line), SH$v_{\rm esc}$H (dotted), SH$v_{\rm
    c}$L (short dashed) and SH$v_{\rm c}$H (long dashed).  Bottom left, modified Maxwellian fits to VL2
  simulation: SHSP (solid line) SIMsh (long dashed), SIMspmed (short
  dashed), SIMsp16 and SIMsp84 (dotted). 
Bottom
 right, dark disc models: SHSP (solid line), DD$\rho$H$\sigma$L
  (dotted), DD$\rho$L$\sigma$L (short dashed), DD$\rho$L$\sigma$H
  (long dashed).
 Note the different scale in
  this and subsequent figures for the dark disc models.}
\label{fig-fv}}

\section{Results}
\label{ddsig} 

\subsection{Differential event rate}

\FIGURE[ht]{
\epsfig{file=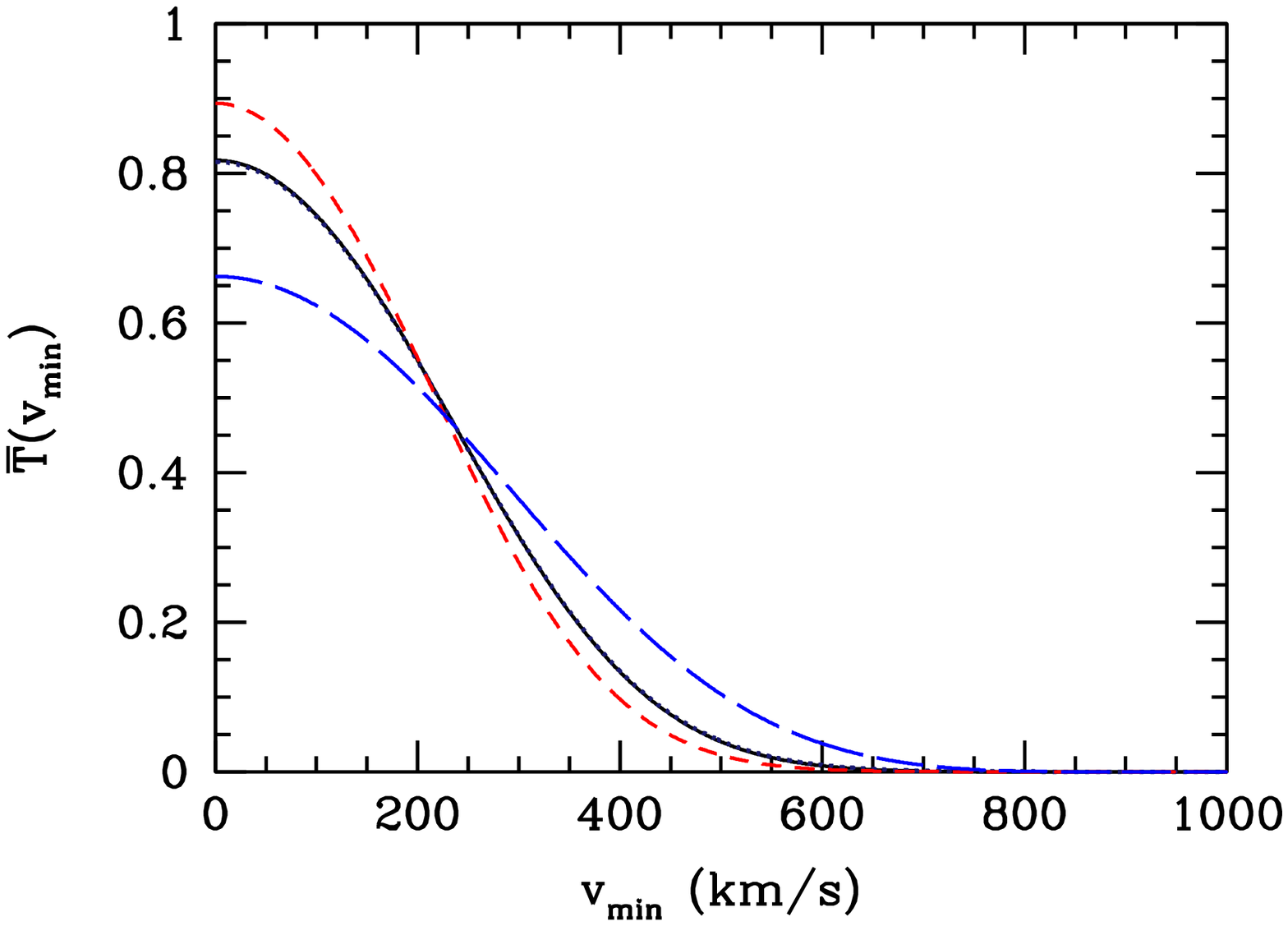,width=7cm}\\
\epsfig{file=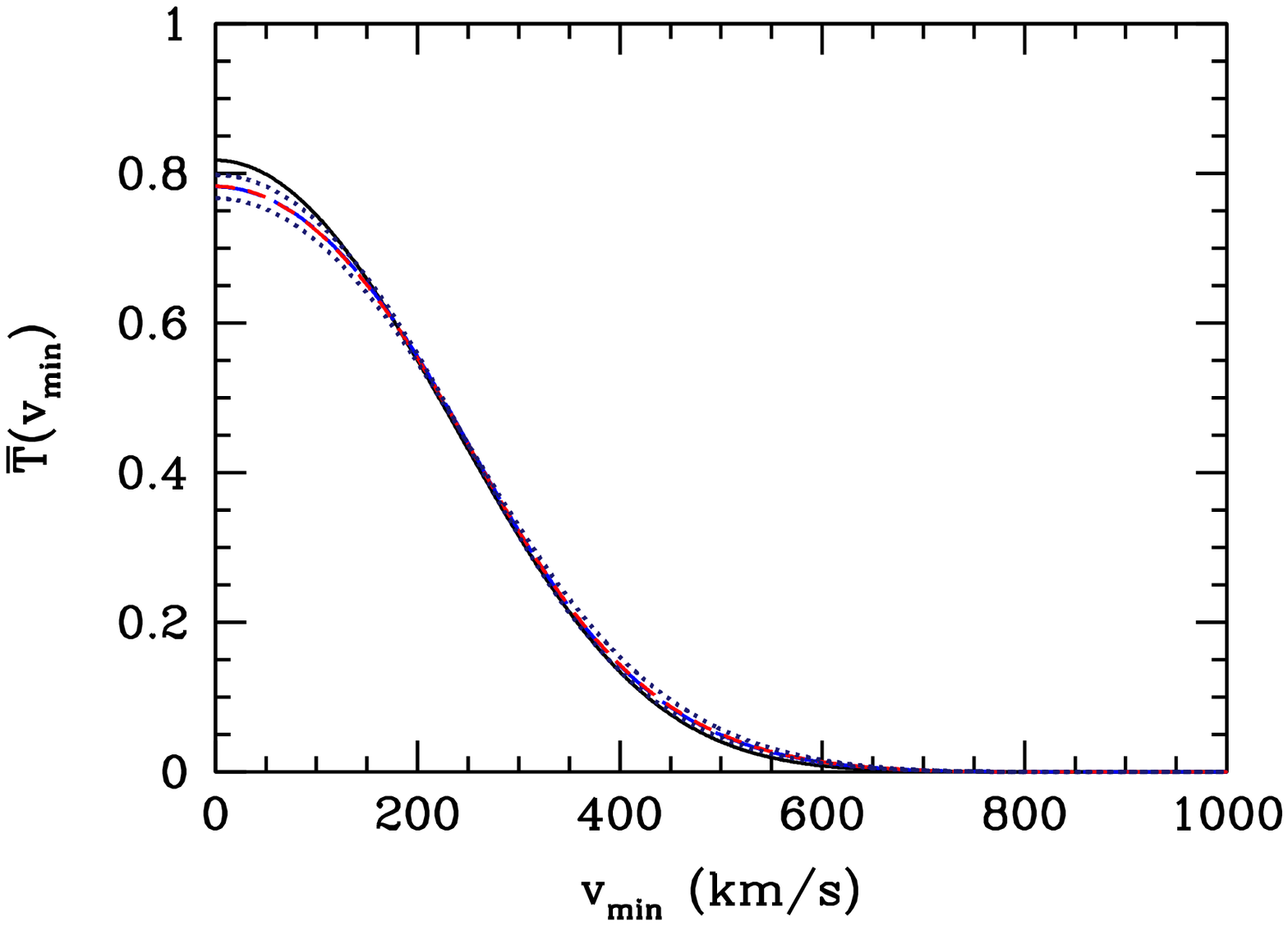,width=7cm}
\epsfig{file=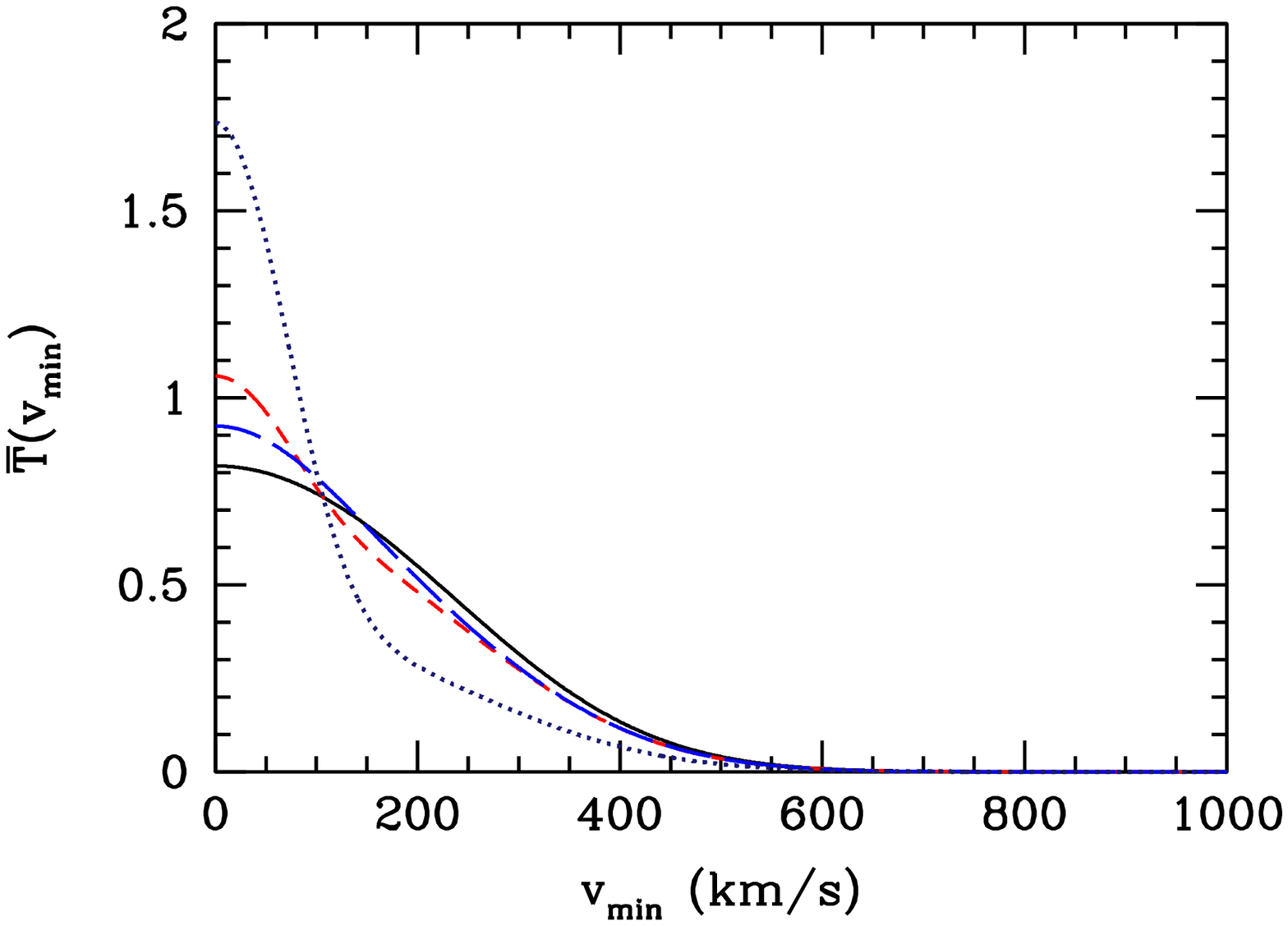,width=7cm}
\caption{The time averaged model independent parameterisation of the differential event rate, $\overline{T}(v_{\rm min})$. Line types as fig. 1.
}
\label{fig-drde}
}

In fig.~\ref{fig-drde} we plot $\overline{T}(v_{\rm min})$, the
time averaged value of the model independent parameterisation of the
differential event rate, eq.(\ref{t}), for each of the benchmark velocity
distributions. For the standard halo model $\overline{T}(v_{\rm min})$
is approximately given by~\cite{ls}
\begin{equation}
\overline{T}(v_{\rm min}) \approx a_{1} \frac{220 \, {\rm km \, s}^{-1}}{v_{\rm c}} \exp{\left( - a_{2}\frac{v_{\rm min}^2}{v_{\rm c}^2} \right)} \,,
\end{equation}
where $a_{1}$ and $a_{2}$ are constants of order unity i.e. increasing
$v_{\rm c}$ decreases both the overall normalisation and the rate at
which $\overline{T}(v_{\rm min})$ decreases with increasing $v_{\rm
  min}$.  Varying the escape speed has a negligible effect, apart from
at large $v_{\rm min}$.  With a DD $\overline{T}(0)$ is larger, and
the initial decrease in $\overline{T}(v_{\rm min})$ as $v_{\rm min}$ is increased
is more rapid. The smaller the DD density and the closer the speed
dispersion to that of the standard halo, the smaller the difference
from the standard halo. For the two models with a low DD
density, the change in $\overline{T}(v_{\rm min})$ is relatively small for $v_{\rm min}
\gtrsim 100 \, {\rm km \, s}^{-1}$. Consequently in these cases there
will only be a significant change in the mean differential event rate
if the WIMP mass and/or experimental energy threshold are sufficiently
low.  For the modified Maxwellian fits to the VL2 simulation data
the shape of $\overline{T}(v_{\rm min})$ is qualitatively similar to that
of the standard halo model, but
$\overline{T}(0)$ is smaller and the fall
off of $\overline{T}(v_{\rm min})$ with increasing $v_{\rm min}$ is slower. The
shell/median sphere fit are similar and the difference between them
and the standard halo is similar to the spread in the sphere fits. The
differences are small, however, compared with those from the
uncertainty in the value of $v_{\rm c}$. We have checked that using
the tabulated data from the VL2 simulation from Ref.~\cite{kuhlenweb} 
produces a similar uncertainty in $\overline{T}(v_{\rm min})$ as the spread
in the sphere fits.

\FIGURE[ht]{
\epsfig{file=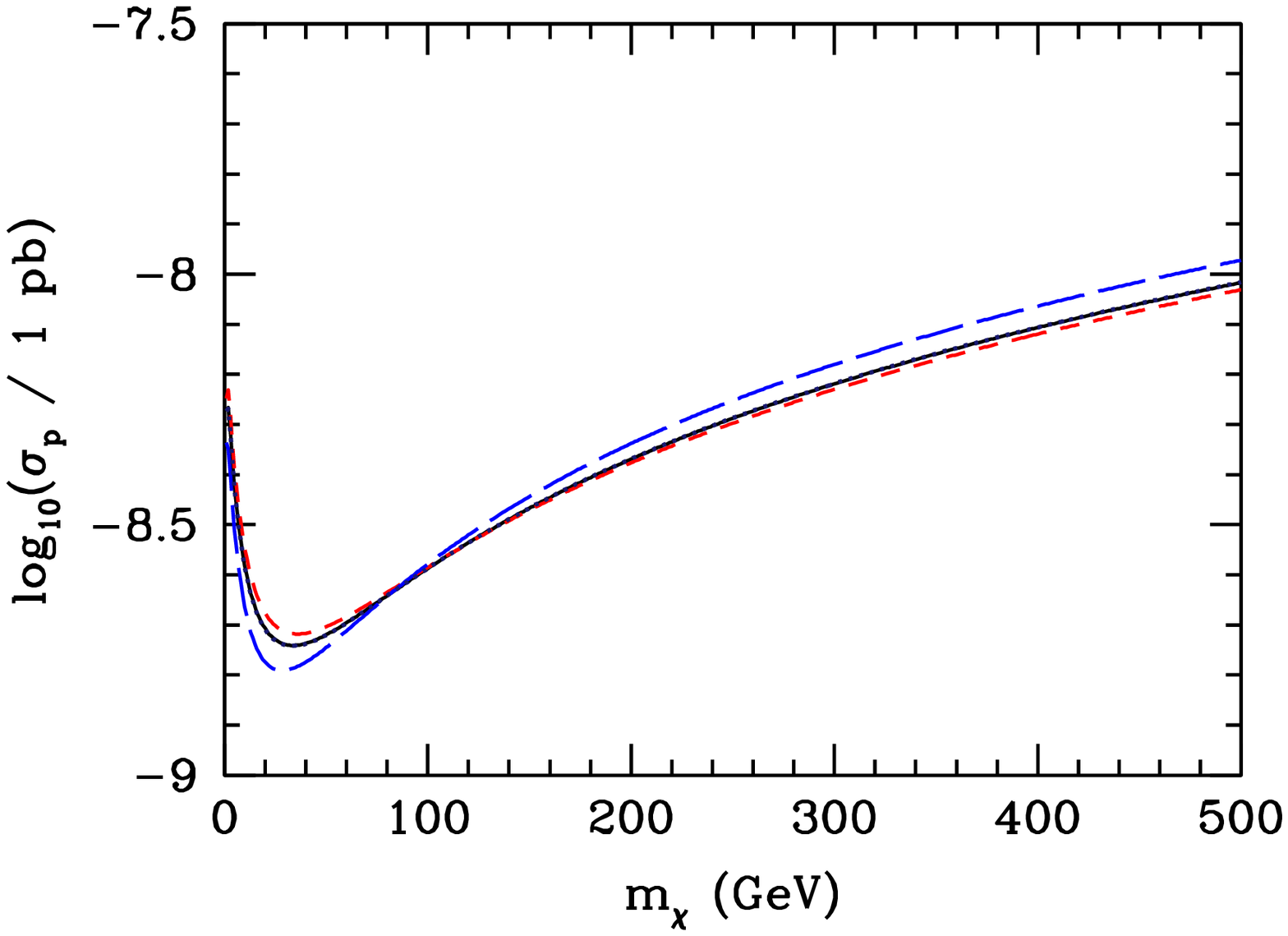,width=7cm}\\
\epsfig{file=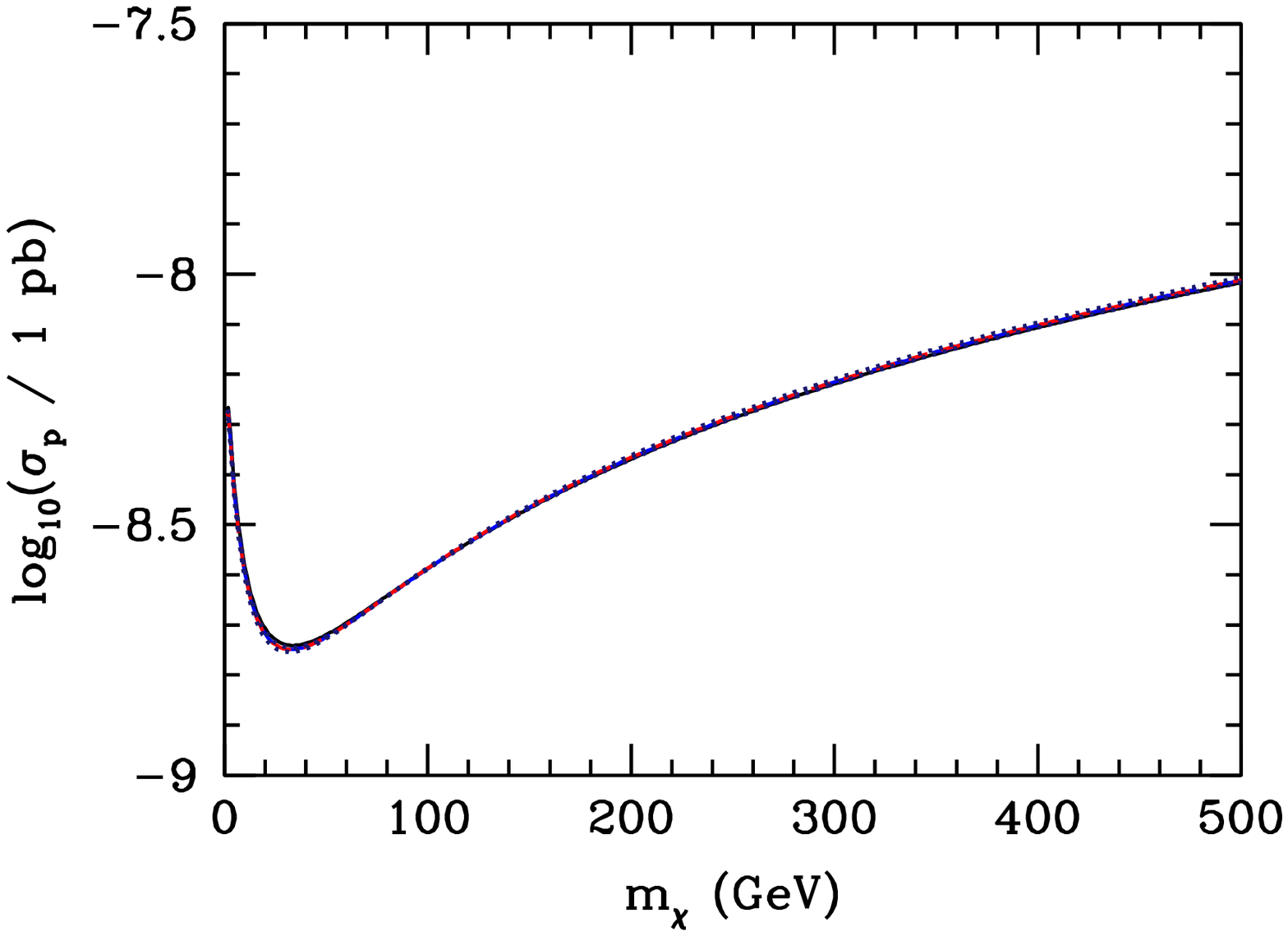,width=7cm}
\epsfig{file=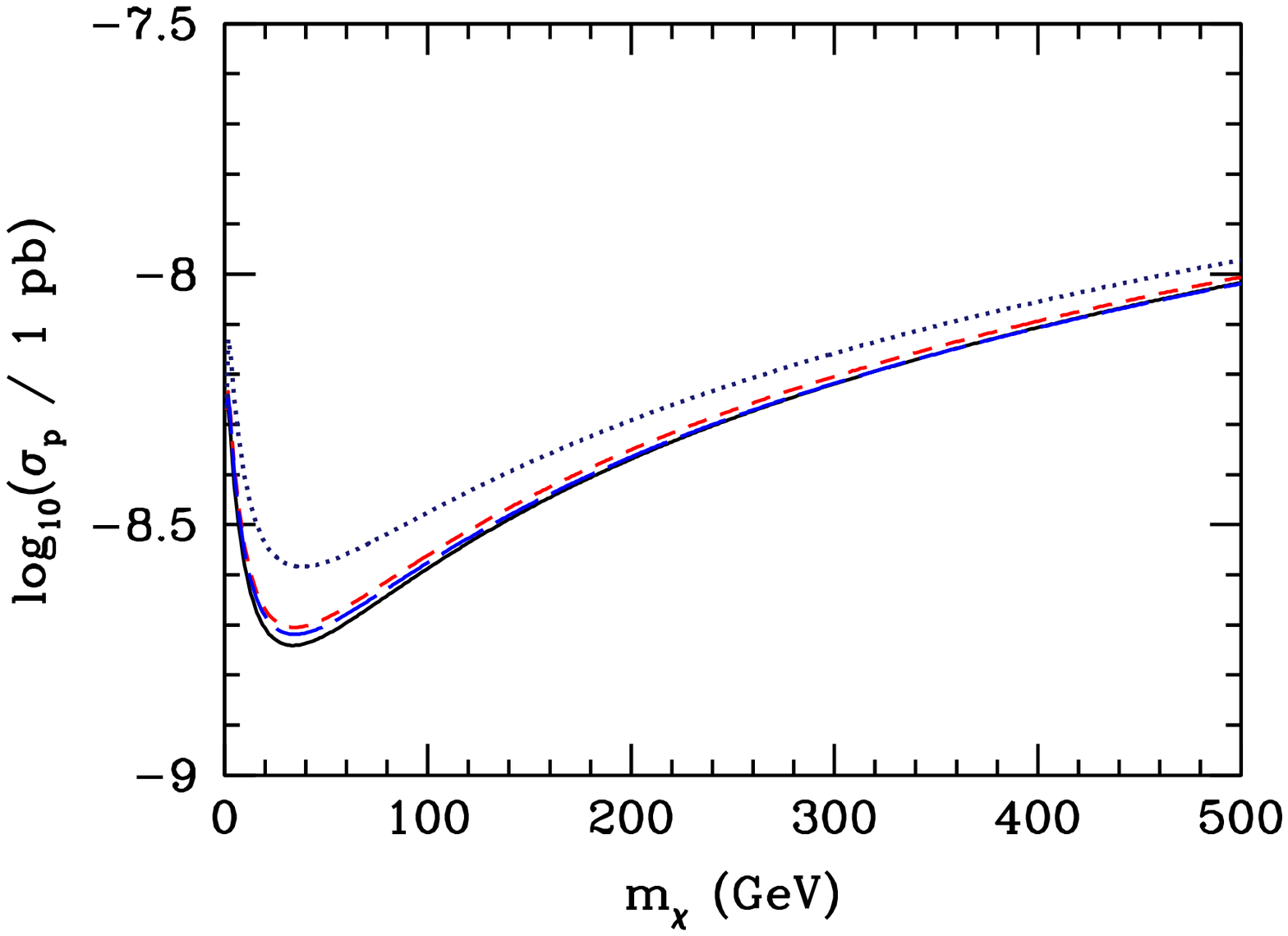,width=7cm}
\caption{Exclusion limits for an ideal experiment with a Ge target and
  an exposure ${\cal E} =  10^{3} \, {\rm kg \,  day}$ which detects zero events. Line types as fig. 1.
}
\label{fig-exclude}
}

In the absence of a positive signal experiments place exclusion limits
on the WIMP mass and cross-section, which depend on the WIMP
distribution~\cite{greenexclude,mccabe}.  To illustrate the effect of
the uncertainty in the WIMP velocity distribution on exclusion limits
we calculate exclusion limits for an ideal experiment (perfect energy
resolution, perfect efficiency, zero background and zero energy
threshold) using a Ge target with an exposure ${\cal E} = 10^{3} \,
{\rm kg \, day}$. For each mass we find the cross-section for which
the expected number of events is equal to three (the $95\%$ upper
confidence limit when zero events are observed)~\cite{pdg}.  The
resulting exclusion limits are shown in fig.~\ref{fig-exclude} for each
velocity distribution.  The total WIMP flux is proportional to the
mean WIMP speed, which for the standard halo model is proportional to
$v_{\rm c}$. Therefore the naive expectation is that increasing
$v_{\rm c}$ increases the total event rate and hence leads to a
tighter constraint on $\sigma_{\rm p}$. As can be seen in
fig.~\ref{fig-exclude} this is the case for small $m_{\chi}$.  For
larger $m_{\chi}$ the, energy dependent, suppression of the event rate
by the form factor, means that when $v_{\rm c}$ is increased the total
event rate in fact decreases and the limit on $\sigma_{\rm p}$ becomes
weaker (see also Ref.~\cite{savagevc}). Increasing the energy
threshold weakens the constraints, and increases the mass at which the
transition occurs. Changing the escape speed only affects the
exclusion limits significantly for light ($m_{\chi} < {\cal O} (10 \, {\rm GeV})$) WIMPs,
see Ref.~\cite{mccabe}. For a high density DD the increase in
$\overline{T}(v_{\rm min})$ for small $v_{\rm min}$ means that the
exclusion limit is a factor of a few weaker for all $m_{\chi}$. For
the low density  DD models and the simulation fits the change in
the exclusion limit is, as expected, relatively small.

Energy dependent efficiency and/or background
events will change the shape of the exclusion limits, see
e.g. Ref.~\cite{greenexclude,mccabe}.  As discussed above the
uncertainty in the local density translates directly into an uncertainty
in $\sigma_{\rm p}$ which is the same for all experiments. Note that,
as pointed out in Ref.~\cite{mccabe}, the values of $\rho_{\chi}$ and $v_{\rm
  c}$ are correlated.

\FIGURE[ht]{
\epsfig{file=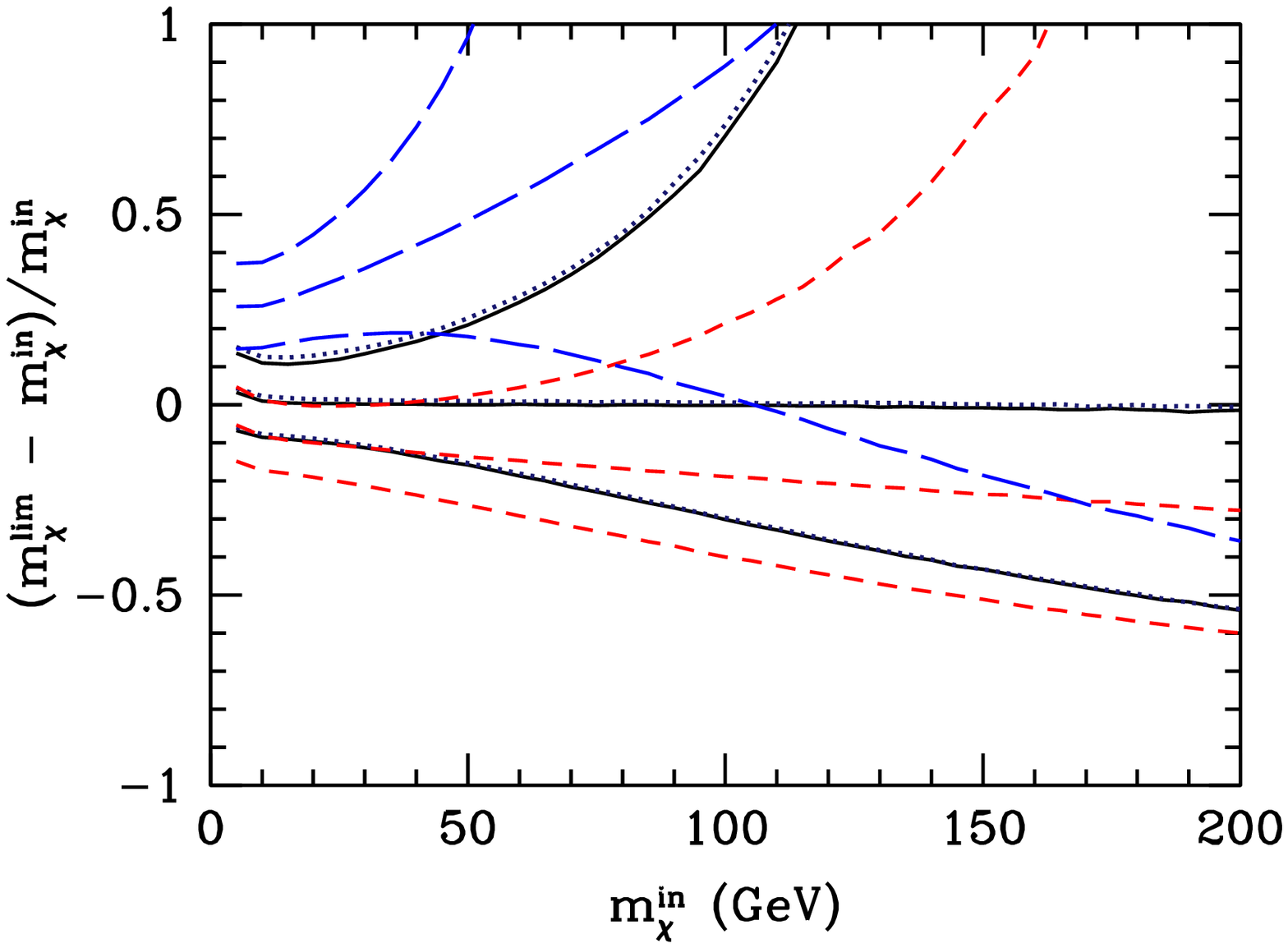,width=7cm}\\
\epsfig{file=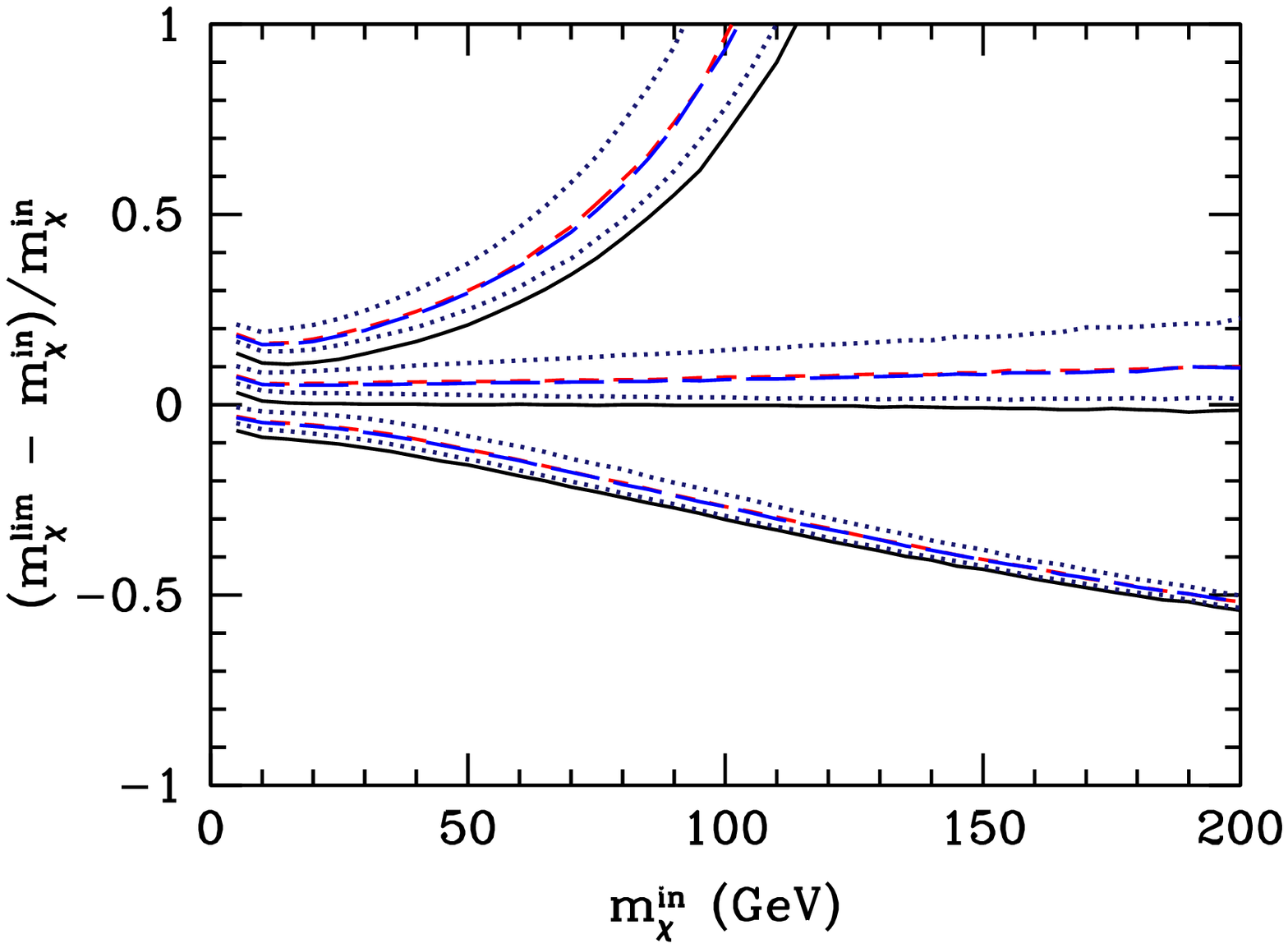,width=7cm}
\epsfig{file=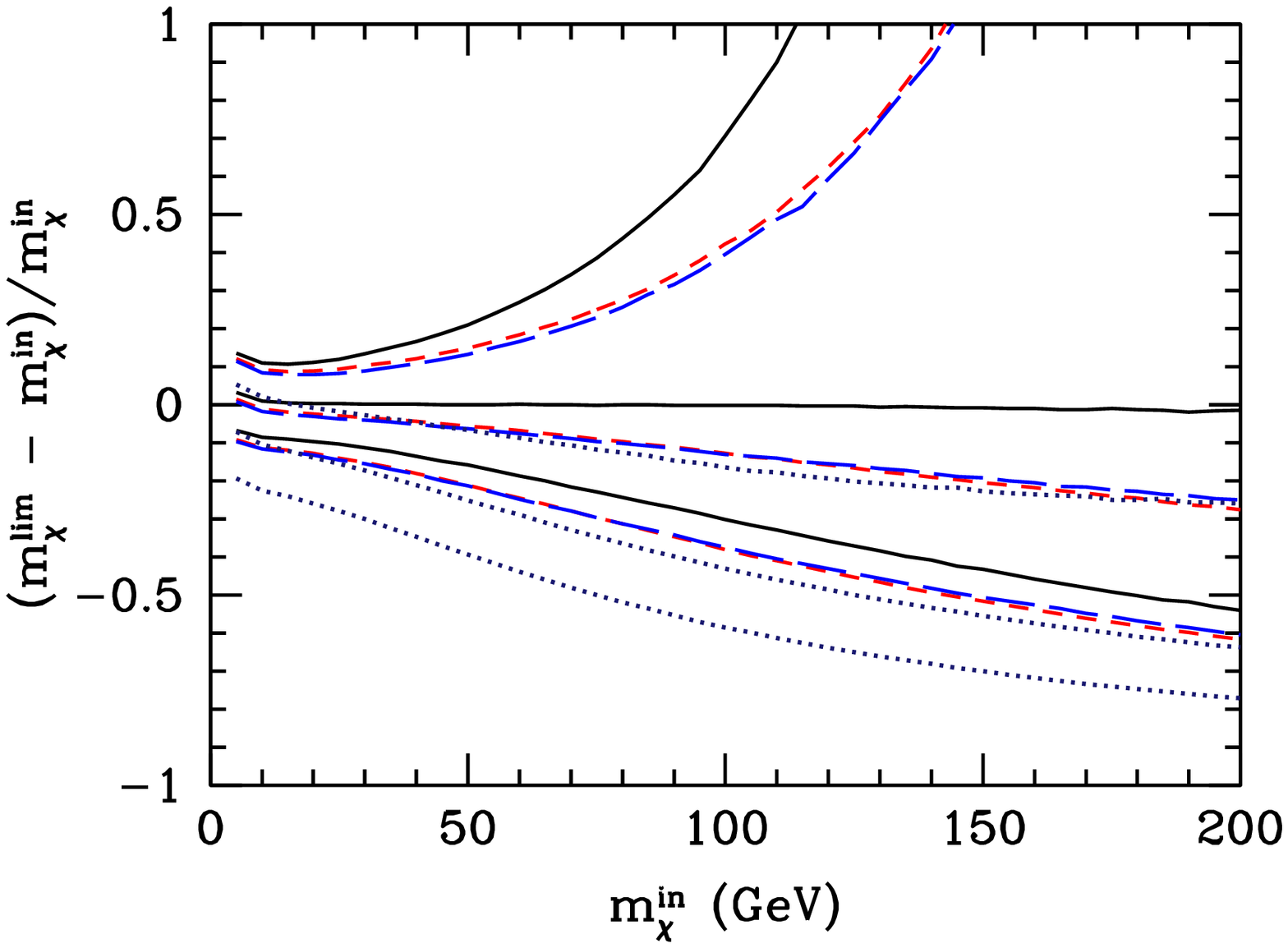,width=7cm}
\caption{Fractional mass limits for an ideal Ge detector and an exposure
  ${\cal E} = 3 \times 10^{4} \, {\rm kg \, day}$ calculated assuming
  the standard halo model with standard parameters. Line types as
  fig. 1.  }
\label{fig-mass}
}

Once events are detected the shape of the energy spectrum can be used
to measure the WIMP mass~\cite{ls,bk,brown,mpap1,sd,mpap2}.
Uncertainty in the velocity distribution leads to a systematic
uncertainty in the determination of the WIMP mass~\cite{mpap1}.  We
use the method used in Refs.~\cite{mpap1,mpap2} to calculate the
limits on the WIMP mass which would be obtained for each velocity
distribution, if the data were analysed assuming the standard halo
model with standard parameters. As before we assume an ideal Ge
detector~\footnote{See Ref.~\cite{mpap2} for an exploration of how
  varying the detector capabilities affects the WIMP mass
  determination.} a WIMP cross-section, $\sigma_{\rm p}= 10^{-8} \,
{\rm pb} $ and an exposure of ${\cal E} = 3 \times 10^{4} \, {\rm kg \,
  day}$.  We estimate the WIMP mass and cross-section by maximising
the extended likelihood function, e.g. Ref.~\cite{cowan}:
\begin{equation} 
  L= \frac{\lambda^{N_{\rm expt}} \exp{(-\lambda)}}{N_{\rm expt}!}
    \Pi_{i=1}^{N_{\rm expt}} f(E_{\rm i}) \,.
\end{equation}
Here $N_{\rm expt}$ is the number of events observed, $E_{\rm i} \,
(i=1,..., N_{\rm expt})$ are the energies of the events observed,
$f(E)$ is the normalised differential event rate and $\lambda = {\cal
  E} \int_{0}^{\infty} ({\rm d} R/{\rm d} E) \, {\rm d} E$ is the mean
number of events.  We calculate the probability distribution of the
maximum likelihood estimator of the WIMP mass, for
each input WIMP mass, by simulating $10^{4}$ experiments and finding
the  $2.5\%$, $50\%$ and $97.5 \%$ percentiles, $m_{\chi}^{\rm lim}$,
of the mass distribution.

The fractional mass limits, $(m_{\chi}^{\rm lim}-m_{\chi}^{\rm in})/m_{\chi}^{\rm in}$
are plotted in  fig.~\ref{fig-mass}  as a
function of the input WIMP mass, $m_{\chi}^{\rm in}$.
As discussed in Refs.~\cite{brown,mpap1,mpap2}, assuming an erroneous value of
$v_{\rm c}$ leads to a systematic error in the mass determination
which increases with increasing $m_{\chi}$:
\begin{equation}
\frac{\Delta m_{\chi}}{m_{\chi}} \sim - \left[ 1 + \left(\frac{m_{\chi}}{m_{\rm
    A}}\right)\right] \frac{\Delta v_{\rm c}}{v_{\rm c}} \,.
 \end{equation}
A dark disc has a similar effect. There is a population of WIMPs with
lower speeds than assumed, and hence the WIMP mass is systematically
underestimated. The systematic underestimate is substantial ($\sim
10-50\%$, increasing with increasing $m_{\chi}$) if the DD density is large and the speed dispersion is significantly
different from that of the halo. For the two models with a low DD
density, the systematic underestimate is smaller, $\sim 5-20 \%$.
The larger width of the modified Maxwellian speed distribution leads
to a systematic overestimate of the WIMP mass in the range $\sim 2-10
\%$, increasing weakly with increasing WIMP mass.

\subsection{Annual modulation}

\FIGURE[ht]{
\epsfig{file=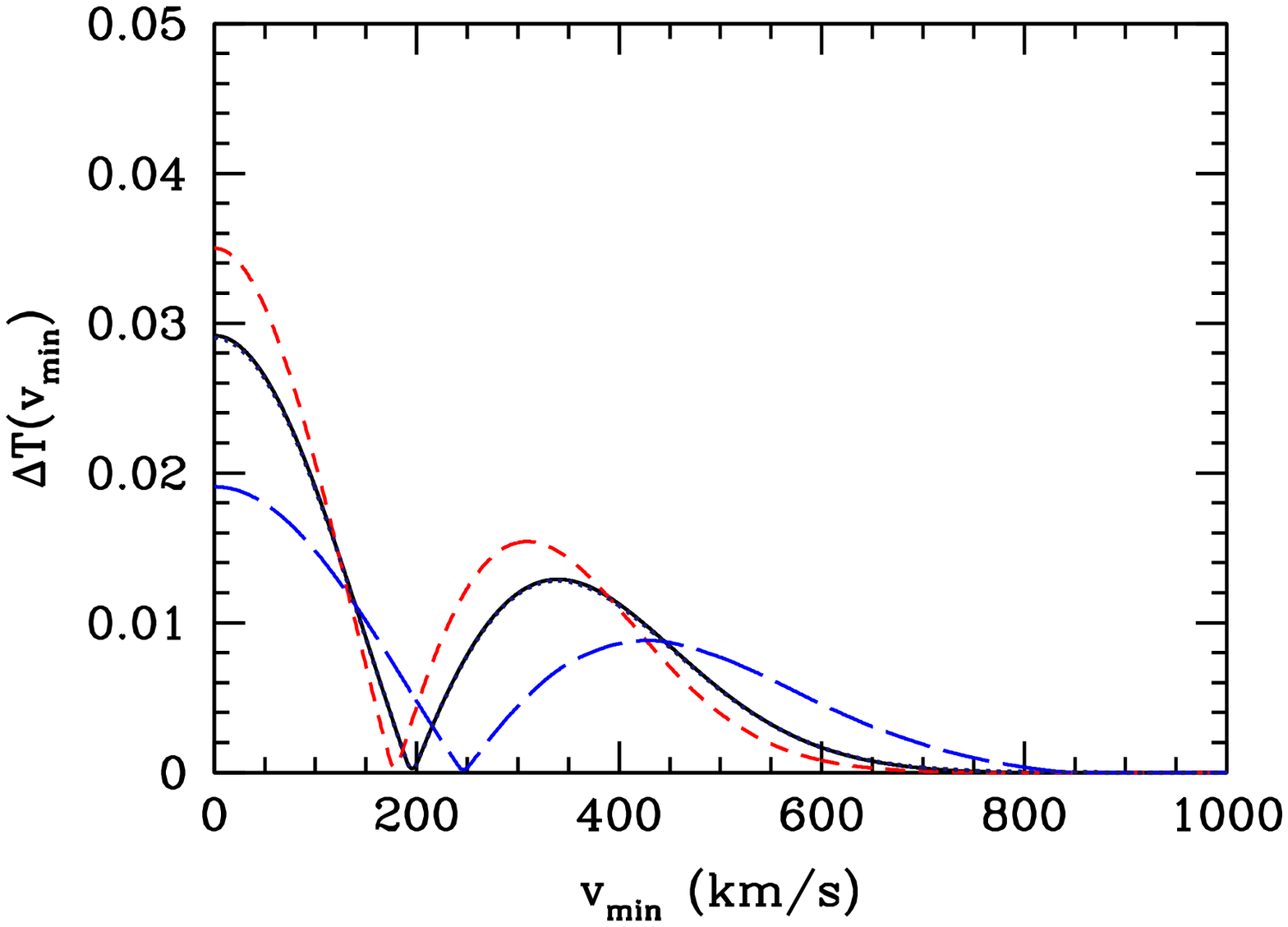,width=7.0cm}\\
\epsfig{file=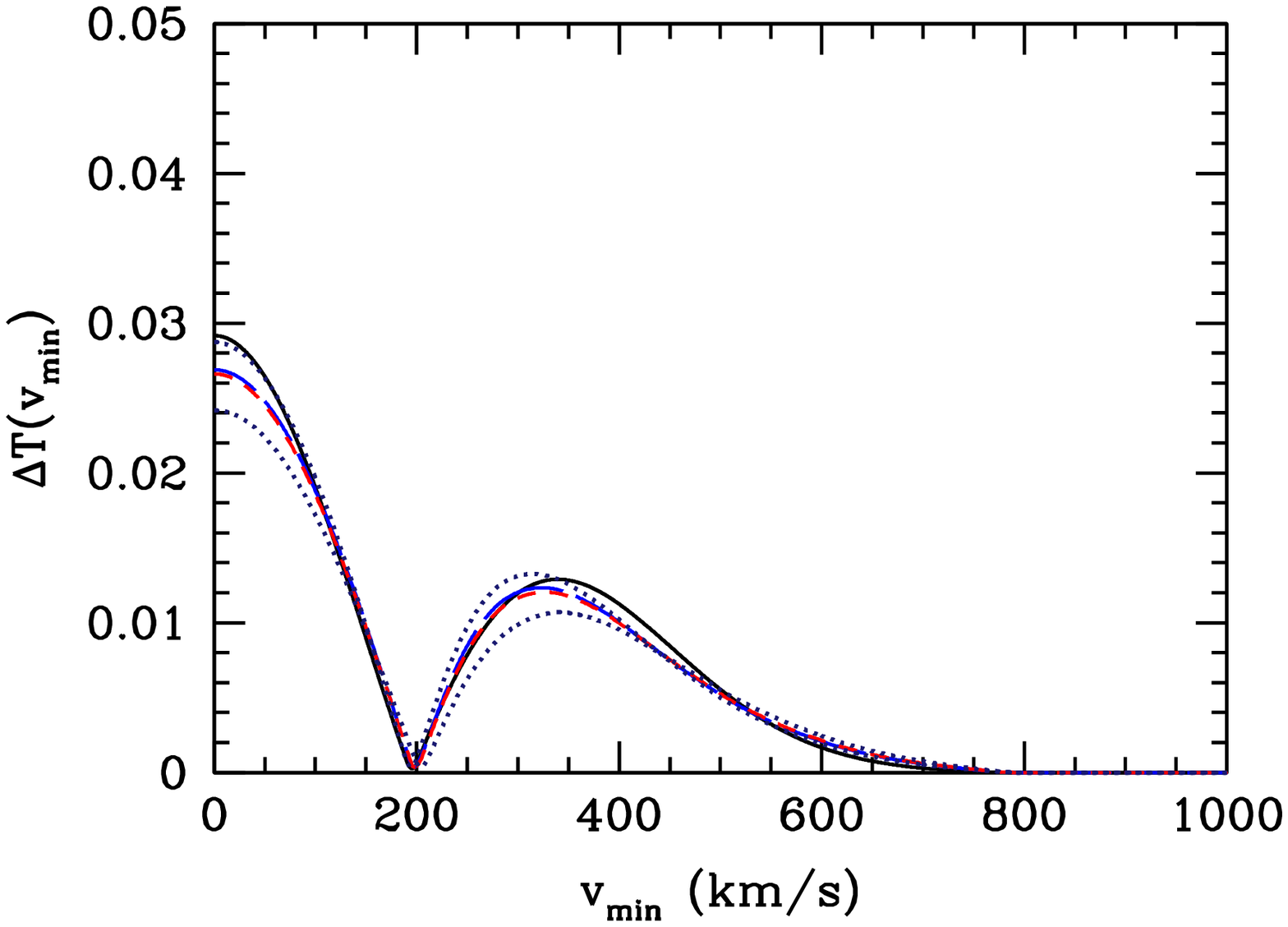,width=7.0cm}
\epsfig{file=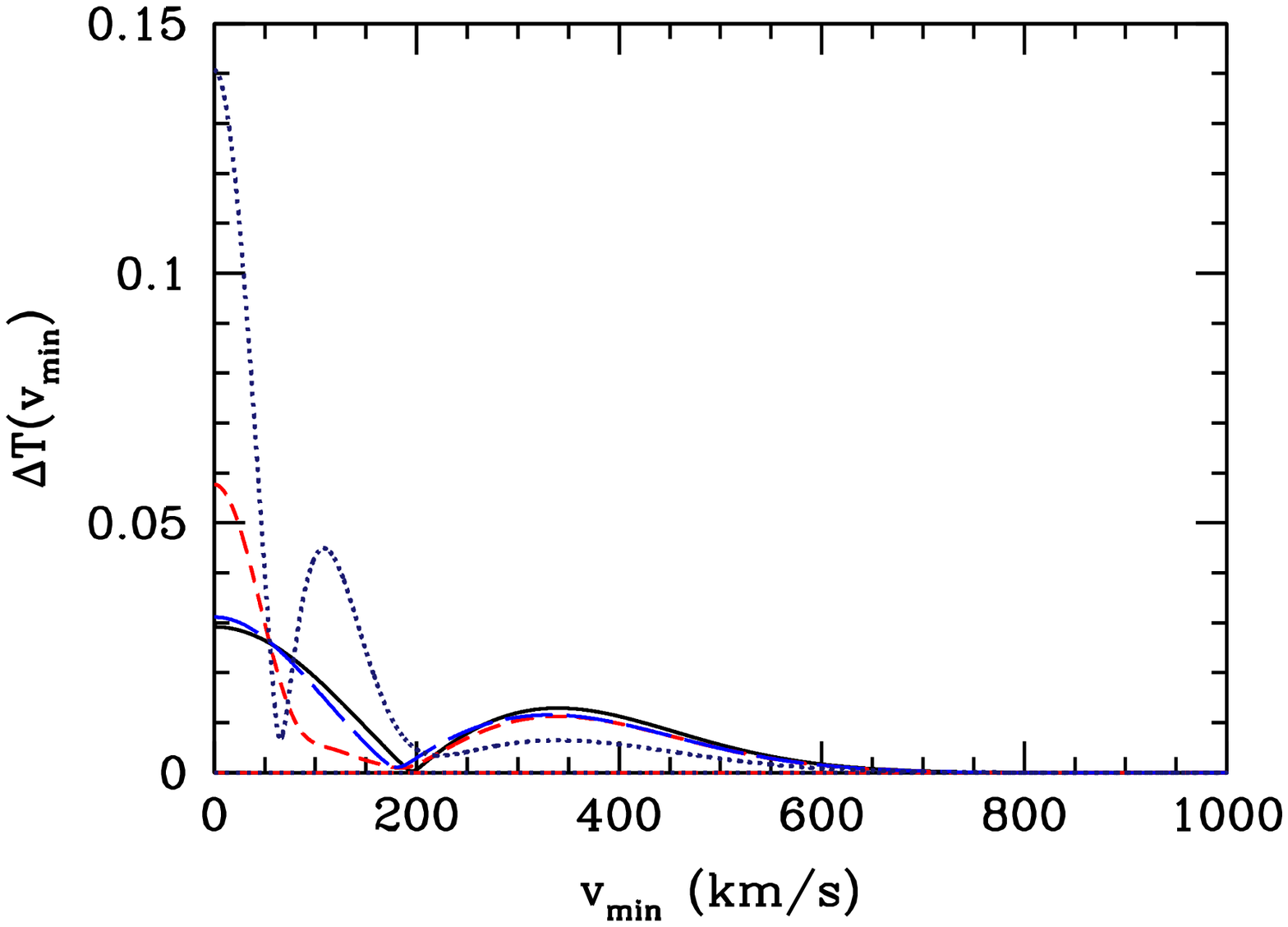,width=7.0cm}
\caption{Amplitude of the annual modulation of the model independent parameterisation of the differential
event rate, $\Delta T(v_{\rm min})$.
 Line types as fig. 1.
}
\label{fig-amp}
}

We consider the amplitude of the modulation in the model independent
parameterisation of the differential event rate, $\Delta T(v_{\rm
  min})={\rm max} [T(v_{\rm min}, t) - \overline{T}(v_{\rm
  min})]$,~\footnote{Some papers, e.g. Ref.~\cite{kuhlen}, consider the
  fractional annual modulation, which is largest for large $v_{\rm
    min}$.}  and
the date $t_{\rm p}$ on which this maximum occurs.  For small $v_{\rm
  min}$, the maximum event rate occurs in Winter~\cite{pss}. As
$v_{\rm min}$ is increased, $\Delta T(v_{\rm min})$ initially decreases to zero
at which point the phase of the annual modulation changes rapidly and the
maximum occurs in Summer. As $v_{\rm min}$ is increased further the
amplitude increases one more to a local maximum (which we refer to as
the Summer maximum), before decreasing again and tending to zero
~\cite{pss,lewis}.

In figs.~\ref{fig-amp} and \ref{fig-amtp} we plot $\Delta T(v_{\rm
  min})$ and the day of the year on which the maximum occurs, $t_{\rm p}$, respectively,
for each of the benchmark velocity distributions.  We do not include a
plot of $t_{\rm p}$ for the standard halo model as the low and high
$v_{\rm min}$ values of $t_{\rm p}$ only change by $\sim 1$ day as $v_{\rm c}$ is varied.

 For the standard halo model, as $v_{\rm c}$ is
increased the most likely velocity and the width of the velocity
distribution both increase. Consequently $\Delta T(0)$ and the Summer
maximum of $\Delta T(v_{\rm min})$ both decrease, the value of $v_{\rm
  min}$ at which the Summer maximum occurs increases
and the decline in the amplitude for large $v_{\rm min}$ is less
rapid. The value of $v_{\rm min}$ at which the maximum switches from
Winter to Summer increases.

For a pure DD (no halo) the behaviour would be qualitatively
similar to the standard halo model. Due to the speed distribution
peaking at a smaller speed, and having smaller dispersion, $\Delta
T(0)$ and the Summer maximum would be larger, and the phase change
would happen at smaller $v_{\rm min}$, $\sim 60 \, {\rm km \,
  s}^{-1}$. When a DD is added to the standard halo the net
effect is more complicated. For DD$\rho$H$\sigma$L there are two local
Summer maxima in the amplitude, a large one at $v_{\rm min} \sim 110
\, {\rm km \, s}^{-1}$ from the DD and a smaller one from the
halo in the usual position, $v_{\rm min} \sim 330 \, {\rm km \,
  s}^{-1}$.  In between $80\, {\rm km \, s}^{-1} \lesssim v_{\rm min}
\lesssim 200 \, {\rm km \, s}^{-1}$ the maximum occurs at $t_{\rm p}
\sim 125$ days.  For DD$\rho$L$\sigma$L the DD density is not
high enough to produce a second Summer maximum, however is does lead
to a significant variation of $t_{\rm p}$ for $80\, {\rm km \, s}^{-1}
\lesssim v_{\rm min} \lesssim 200 \, {\rm km \, s}^{-1}$. This is
because for these values of $v_{\rm min} $ the contributions from the
DD and the halo are out of phase, and partly cancel. For
DD$\rho$L$\sigma$H the change from the standard halo model is fairly
small.

For the modified Maxwellian distributions the qualitative behaviour of
$\Delta T(v_{\rm min})$ is the same as for the standard halo model;
$\Delta T(0)$ is smaller, the Summer maximum occurs at a smaller
$v_{\rm min}$, but its amplitude can be either smaller or larger. The
changes in the amplitude of the modulation are larger than those in
the mean differential event rate. However, the changes from the
uncertainty in $v_{\rm c}$ are again larger than those from the
uncertainty in the shape of the speed distribution (unless there is a
dark disc with small speed dispersion). The change in $t_{\rm p}$ from
Winter to Summer happens more gradually, and for large $v_{\rm min}$
$t_{\rm p}$ is typically a few days smaller. For both the SHSP and the
modified Maxwellian fits $\Delta T (\sim  200
\, {\rm km \, s}^{-1} ) \sim 0$ and hence the corresponding value of
$t_{\rm p}$ is hard to calculate accurately, and not particularly
physically meaningful.

\FIGURE[ht]{
\epsfig{file=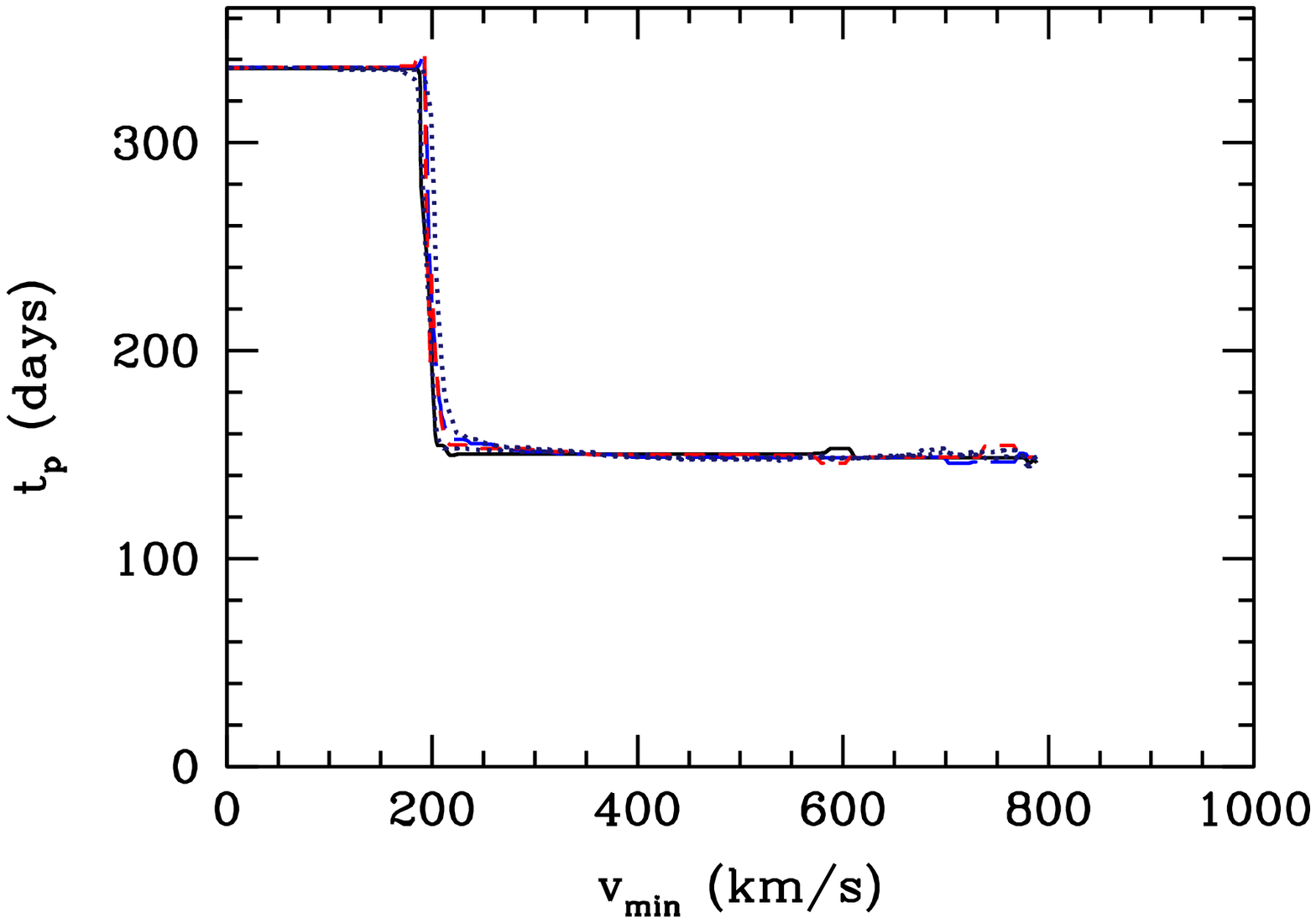,width=7.0cm}
\epsfig{file=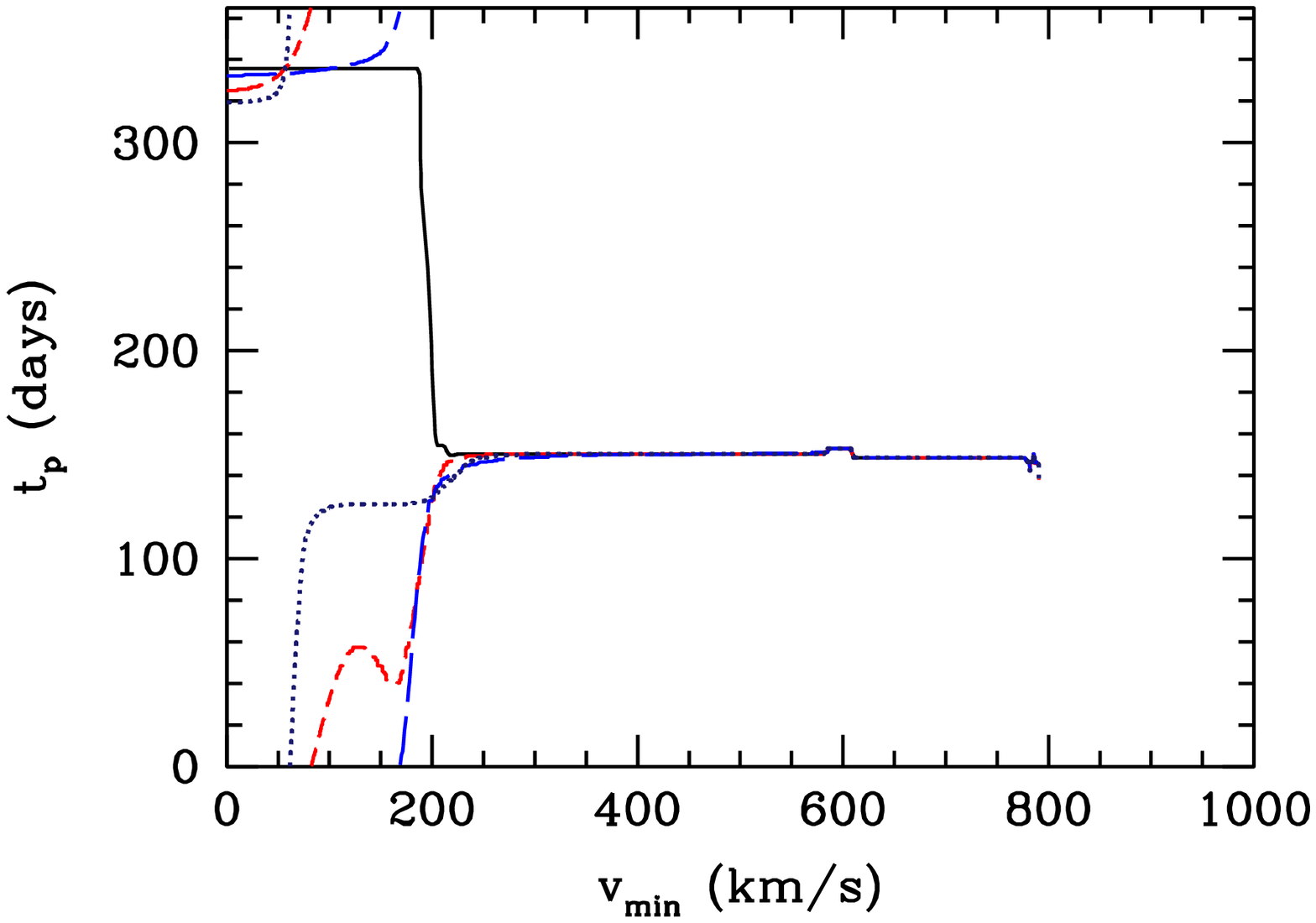,width=7.0cm}
\caption{Day of the year, $t_{\rm p}$, on which the maximum of the differential
  event rate, $\Delta T(v_{\rm
  min})={\rm max} [T(v_{\rm min}, t) - \overline{T}(v_{\rm
  min})]$, occurs.  Line types as Fig. 1 for the modified Maxwellian
fits to the  simulation data (left panel) and
  dark disc models (right panel).}
\label{fig-amtp}
}

\subsection{Direction dependence}

\FIGURE[ht]{
\epsfig{file=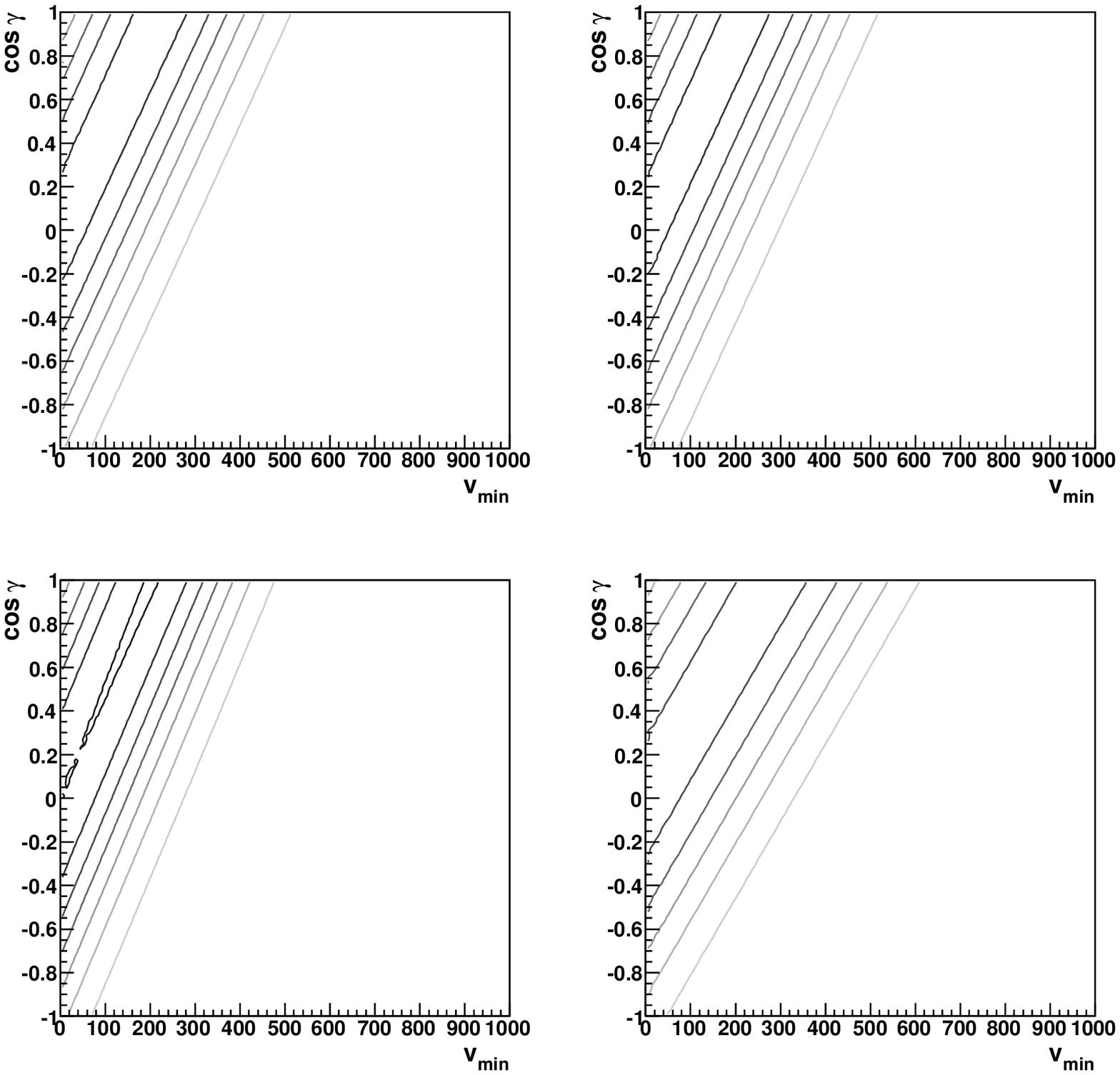, width=12.0cm} 
\caption{Contour plots of the model independent parameterisation of the direction
dependence, ${\cal T}(v_{\rm min}, \cos{\gamma})$, for the standard
halo model. Top row: SHSP and SH$v_{\rm esc}$H, bottom row: SH$v_{\rm c}$L and 
SH$v_{\rm c}$H. Contours have spacing 0.5 between 0 and 3.5.}
\label{fig-dd1}
}

\FIGURE[ht]{
\epsfig{file=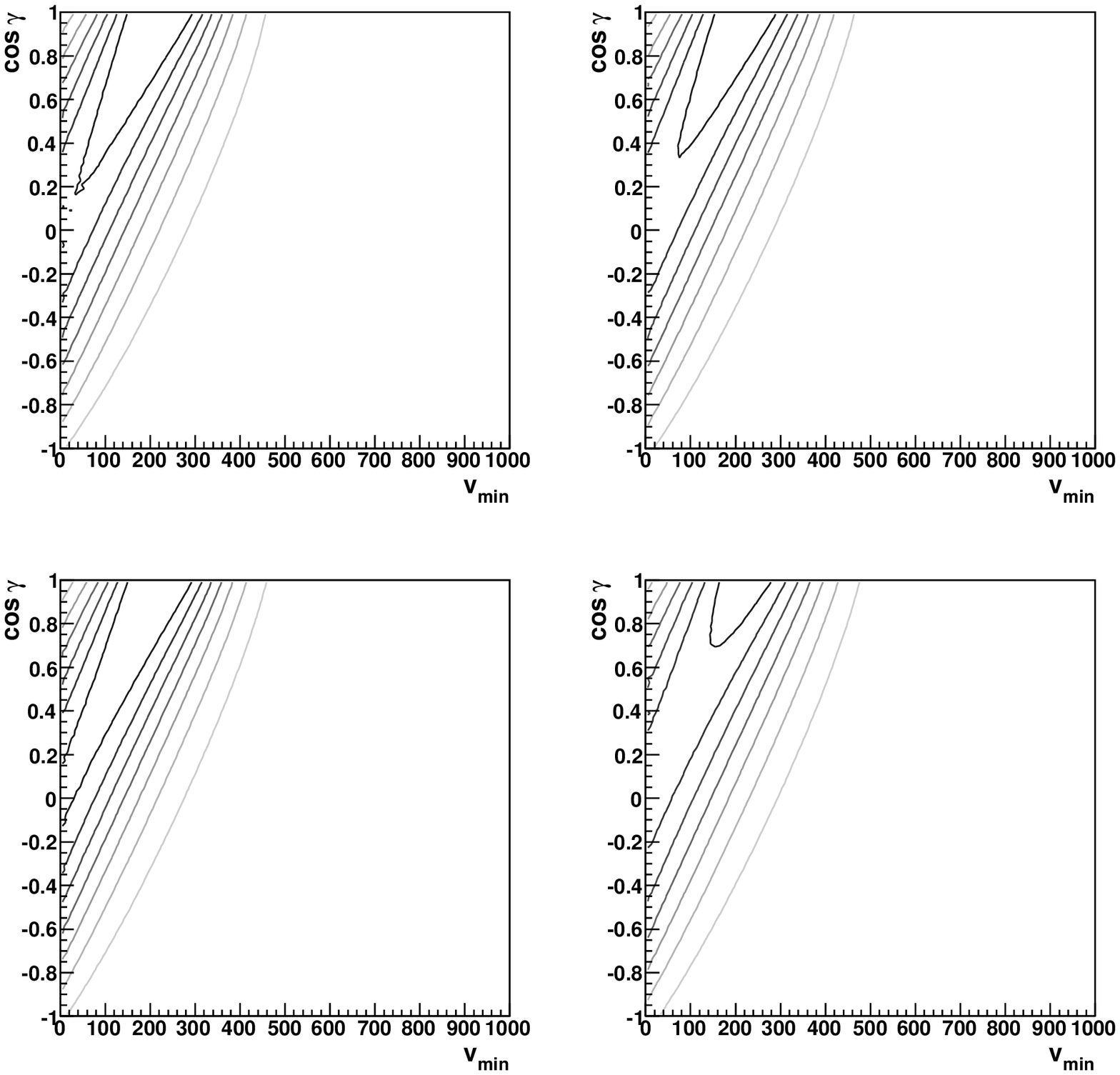, width=12.0cm} 
\caption{As fig.~\ref{fig-dd1}  for the modified Maxwellian fits to the simulation data.
Top row: SIMsh and SIMspmed, bottom row: SIMsp16 and SIMsp84. The contour
levels are the same as in fig.~\ref{fig-dd1}. }
\label{fig-dd2}
}

\FIGURE[ht]{
\epsfig{file=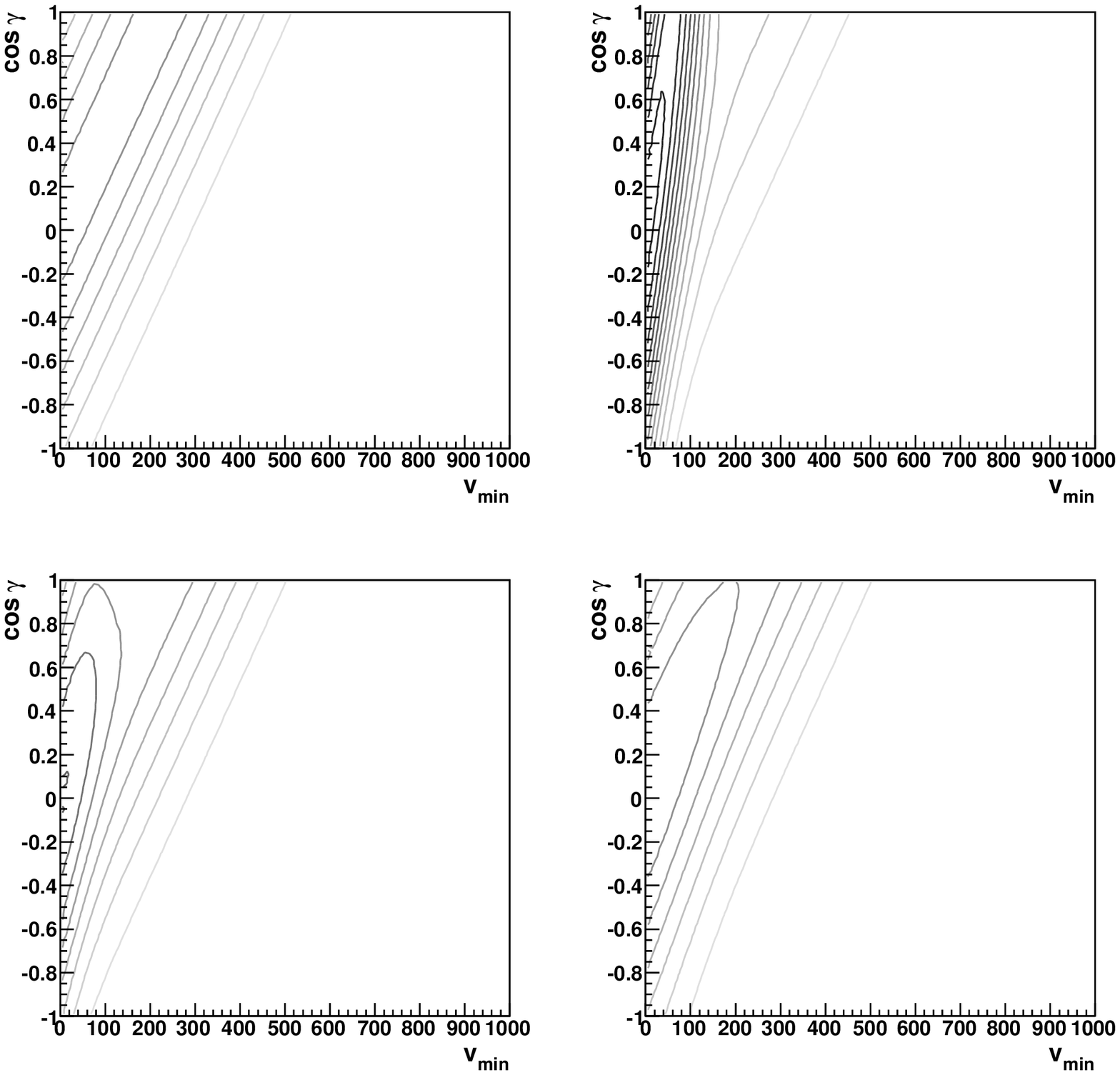, width=12.0cm} 
\caption{As fig.~\ref{fig-dd1}  for the dark disc models (and, for
  comparison, the standard halo model with standard parameters).
Top row: SHSP and
DD$\rho$H$\sigma$L, bottom row: DD$\rho$L$\sigma$L and DD$\rho$L$\sigma$H.
For clarity these plots have a different scale to figs.~\ref{fig-dd1}
and \ref{fig-dd2}.  Contours have spacing 0.5 between 0 and 6.5. }
\label{fig-dd3}
}

The model independent parameterisation of the direction
dependence, ${\cal T}(v_{\rm min}, \cos{\gamma})$ as defined in eq.~(\ref{tdir}), 
is shown in figs.~\ref{fig-dd1}, \ref{fig-dd2} and \ref{fig-dd3} for
the standard halo model, the modified Maxwellian fits to the simulation data and the dark
disc models respectively. 
For the standard halo
model~\cite{spergel} 
\begin{equation}
{\cal T}(v_{\rm min}, \cos{\gamma}) \approx 2\sqrt{\pi} \exp{ \left[ -
    \left( \frac{ (v_{\rm e}^{\rm orb, p}  + v_{\odot}) \cos{\gamma} -
        v_{\rm min} }{v_{\rm c}} \right)^2 \right]} \,, 
\end{equation}
where $v_{\rm e}^{\rm orb, p}$ is the component of the Earth's
velocity parallel to the direction of Solar motion. This has a maximum
value, for $v_{\rm min}= ( v_{\rm e}^{\rm orb, p}  + v_{\odot}   )
\cos{\gamma}$, of ${\cal T}  \approx 2\sqrt{\pi} \approx 3.5$. As can be seen in
fig.~\ref{fig-dd1}, for the standard halo model ${\cal T}(v_{\rm min},
\cos{\gamma})$ is constant for fixed $v_{\rm min}/\cos{\gamma}$. This
is not the case for the other velocity distributions. For the
anisotropic modified Maxwellian distribution, around the peak of the
${\cal T}(v_{\rm min}, \cos{\gamma})$ distribution as $v_{\rm min}$ is
decreased, with $v_{\rm min}/\cos{\gamma}$ fixed, the value of ${\cal
  T}$ decreases, while the contours of low fixed ${\cal T}$ are
convex.  For the DD models the additional population of low speed
WIMPs means that ${\cal T}$ peaks at smaller $v_{\rm min}$ and has a
larger peak value ($\approx 6.5$ for DD$\rho$H$\sigma$L compared with
$\approx 3.5$ for the standard halo). Around the peak of the ${\cal T}(v_{\rm min}, \cos{\gamma})$
distribution in this case as $v_{\rm min}$ is decreased, with $v_{\rm
  min}/\cos{\gamma}$ fixed, the value of ${\cal T}$ decreases. As
before, the smaller the DD density and the closer the speed dispersion
to that of the standard halo, the smaller the difference from the
standard halo.

We use the methods presented in Ref.~\cite{morganpap1} to calculate
the number of events required to determine that the recoil
distribution is not isotropic if $m_{\chi}= 100 \, {\rm GeV}$. Briefly, we use the HADES
code~\cite{bm:thesis} to simulate the recoil direction distribution for a $S$ detector
with energy threshold $20 \, {\rm keV}$~\footnote{It is impossible for a directional detector
to have zero energy threshold, even in principle, as the lengths of the nuclear recoils tend to zero in this limit, and hence their direction is unmeasurable.}.   We
assume that the recoil directions, including their senses, are
reconstructed perfectly in 3d and the background is zero. These are optimistic
assumptions and therefore
our results provide a lower limit on the number of events required by
a real detector. For 3-d data the most powerful test for
rejecting isotropy uses the average of the cosine of the angle between
the direction of solar motion and the recoil direction, $\langle
\cos{\gamma} \rangle$~\cite{morganpap1}. We calculate the probability
distribution of $\langle \cos{\gamma} \rangle$, for a given number of
events $N$, by Monte Carlo generating $10^4$ experiments and find
the number of events required to reject isotropy at $95\%$ confidence
in $95\%$ of experiments, $N_{\rm iso}$. For further details see
Ref.~\cite{morganpap1}.   For all the velocity distributions
considered  $N_{\rm iso}=9$.

The WIMP origin of an anisotropic recoil distribution could be checked
by measuring the median recoil direction~\cite{wimpdirconf}. For a
smooth WIMP distribution the median inverse recoil direction coincides
with the direction of solar motion, modulo statistical
fluctuations. 
The median inverse direction is defined as the direction ${\bf x}_{\rm med} $
which minimises the sum of the arc-lengths between ${\bf x}_{\rm med}$
and the individual inverse recoil directions ${\bf x}_{\rm
  i}$~\cite{fisher:lewis:embleton}. It is found by minimising
\begin{equation}
\label{M}
{\cal M} = \sum_{i=1}^{N} {\rm cos}^{-1} ({\bf x}_{\rm med}.{\bf x}_{\rm i}) \,,
\end{equation}
where $N$ is the number of events. 
We determine the number of events required to confirm the direction of
solar motion as the median 
inverse recoil direction at $95\%$ confidence using the distribution of $\Delta$,
the angle between the median direction and the direction of solar
motion, ${\bf x}_{\odot}$:
\begin{equation}
  \Delta= {\rm cos}^{-1}({\bf x}_{\rm med}.{\bf x}_{\odot}) \,.
\end{equation}
We calculate the probability distribution
of $\Delta$, for a given number of events $N$, by Monte Carlo
generating $10^4$ experiments and find the number of events
required to reject the median direction being random at $95\%$
confidence in $95\%$ of experiments, $N_{\rm med}$. For further
details see Ref.~\cite{wimpdirconf}. The value of $N_{\rm med}$,
is only weakly dependent on the velocity distribution.
For SH$v_{\rm c}$H, $N_{\rm med}=32$ while for all the other velocity distributions
$N_{\rm med}=27$ or $28$.

The limited variation of $N_{\rm iso}$ and $N_{\rm med}$ show that the
directional signals are robust to (plausible) uncertainties in the
velocity distribution. The modified Maxwellian fits to the simulation
data do not include the stochastic features found at high speeds however.
These features can cause the median inverse direction of high energy recoils to
deviate from the direction of solar motion~\cite{kuhlen}. As discussed
in Ref.~\cite{wimpdirconf} the deviation will be small unless the
detector is only sensitive to the high speed tail of the speed
distribution (i.e. if the WIMP mass is small and/or the energy
threshold is high). Studying the median direction of high energy
recoils could in fact allow high speed features to be detected, hence
probing the formation history of the Milky Way.

\section{Summary}
\label{discuss}

Direct detection event rate calculations often assume the standard
halo model, with an isotropic Maxwellian velocity distribution.
It is well known, however, that the energy~\cite{kk,donato,greenexclude,vhh,bruch,vogelsberger,kuhlen,ling1,ling2,mccabe},
time~\cite{br,belli,vergados,ullio,ecz,am1,am2,nata,vhh,bruch,fs,vogelsberger,kuhlen,ling1,ling2} and
direction~\cite{ck1,ck2,morganpap1,kuhlen} dependence of the direct detection
event rate all depend on the local WIMP distribution.  We have updated
these studies in light of recent numerical simulations and
observational measurements of the local circular speed and escape
speed.

The local circular speed has a model dependent relation to the
radial velocity dispersion. Recent determinations,
e.g. Refs~\cite{reid,bovy}, have relatively small statistical errors,
however Ref.~\cite{mcmillan} found values ranging from $v_{\rm c} =
200$ to $280 \, {\rm km \, s}^{-1}$, depending on the model of the MW
assumed.  The most recent determination of the local escape
speed~\cite{smith} finds $v_{esc} = 544 \, {\rm km \, s}^{-1}$,
slightly lower than the historical standard value, $v_{esc} = 650 \,
{\rm km \, s}^{-1}$. Similarly for a given model for the MW it is
possible to determine the local DM density, $\rho_{\chi}$, to $\sim 10\%$
accuracy~\cite{widrow,cu}, however the systematic errors are likely to
be substantially larger~\cite{weber,garbari,salucci,pato}.

High resolution dark matter only simulations of the formation of Milky
Way like dark matter halos, typically find speed distributions which
can be fit with a modified Maxwellian distribution which is broader
than the standard Maxwellian
distribution~\cite{hansen,fs,vogelsberger,kuhlen}.  Some simulations
which include baryonic physics find that late merging sub-halos
are preferentially dragged towards the disc, where they are destroyed
leading to the formation of a co-rotating dark
disc~\cite{read1,read2,ling1}.  The significance of a DD for direct detection
experiments depends on its density and velocity distribution,
which are highly uncertain~\cite{bruch,purcell,ling2}.

We have considered three types of velocity distribution: standard
Maxwellian, modified Maxwellian, standard Maxwellian plus dark
disc. In each case we use a range of parameter values motivated by
recent observations and simulations (as discussed in detail in Sec.~\ref{dist}).\\

{\bf Differential event rate and exclusion limits}

The systematic uncertainty in the local circular speed, $v_{\rm c}$,
leads to a ${\cal O}(10 \%)$ uncertainty in the differential event
rate, and hence exclusion limits. A high density DD also
produces significant changes in these quantities. A low density DD
will only have a significant effect if the WIMP mass and/or energy
threshold are sufficiently low.  The dependence on the detailed shape
of the velocity distribution is small.

The normalisation of the differential event rate is directly
proportional to the product of the local density and the WIMP
cross-section. Therefore the uncertainty in the measurement of the local
density propagates directly into an uncertainty on measurements of, or
constrains on, the cross-section.\\

{\bf Mass determination} 

We studied the systematic errors in
determinations of the WIMP mass which would occur if data from a
future SuperCDMS like experiment is, erroneously, analysed assuming
the standard halo model with standard parameters.
Assuming an incorrect value of $v_{\rm c}$ leads to a systematic error in the mass
determination which increases with increasing
$m_{\chi}$~\cite{brown,mpap1,mpap2}. With a DD there is a
population of WIMPs with lower speeds than assumed and hence the WIMP
mass is underestimated. The size of the systematic error varies, with
increasing $m_{\chi}$, from $10-50\%$ for a high density dark disc and
from $2-10\%$ for a low density dark disc.  The larger width of the
modified Maxwellian distribution leads to a smaller overestimate of $m_{\chi}$.\\

{\bf Annual modulation} 

The annual modulation is more sensitive to
the velocity distribution than the mean differential rate. Changing
the value of $v_{\rm c} $ or a DD with small speed dispersion
can change the amplitude by a factor of order unity, while the
uncertainty in the shape of the halo velocity distribution changes the
amplitude by $\sim {\cal O}(10 \%)$. 

The phase of the modulation depends only weakly on $v_{\rm c}$ and the
shape of the halo velocity distribution. With a small speed dispersion
DD the phase at moderate energies changes by $\sim 10 \, (100)$ days if
the DD density is low (high).  \\

{\bf Direction dependence}

The detailed direction dependence of the event rate is sensitive to
the velocity distribution, however the directional signals are
robust. The number of events required to detect anisotropy does not
change, while the number of events required to demonstrate that the
median inverse recoil direction
coincides with the direction of solar motion varies by of order $10\%$.\\

Even with recent improvements in the resolution of numerical
simulations and observational determinations of the dark matter
parameters, there are still significant uncertainties in the direct
detection signals and WIMP parameter determinations. In particular the
existence of a DD could have a significant effect, depending on
its density and speed dispersion.

Considering a range of, data motivated, benchmark models is an
improvement on simply assuming the `standard halo model'. However it
is still not completely satisfactory. In particular it does not
provide a formal analysis of errors.  Several approaches to
dealing with the impact of astrophysical uncertainties on direct
detection data analysis have recently
been proposed. Strigari and Trotta~\cite{st} have suggested using
astronomical data and a model for the Milky Way mass distribution in a
Monte Carlo Markov Chain analysis of direct detection
data. Peter~\cite{peter} has presented an approach which involves
combining data sets from different direct detection experiments and
jointly constraining a parametrisation of the WIMP speed distribution
and the WIMP parameters (mass and cross-section). These are promising
directions, however in both cases the current implementations assume an 
overly restrictive form for the velocity distribution (a single isotropic
Maxwellian).

\acknowledgments AMG is supported by STFC, is grateful to Ben Morgan, Mike Kuhlen
and Simon Goodwin for useful discussions and acknowledges the use of Ben Morgan's
HADES, directional detection code.

\end{document}